\documentclass[journal]{IEEEtran}

\usepackage{amssymb}
\usepackage{latexsym}
\usepackage{eccvabbrv}
\usepackage{siunitx}
\usepackage{rotating}
\usepackage{url}
\usepackage{booktabs}
\usepackage{color,soul}

\usepackage{amsmath}
\usepackage{mathtools}
\usepackage{enumitem}

\usepackage{dblfloatfix}
\usepackage{tikz}       
\usepackage{nicematrix} 
\usepackage{framed,multirow}

\usepackage{booktabs,arydshln}
\makeatletter
\def\adl@drawiv#1#2#3{%
        \hskip.5\tabcolsep
        \xleaders#3{#2.5\@tempdimb #1{1}#2.5\@tempdimb}%
    #2\z@ plus1fil minus1fil\relax
        \hskip.5\tabcolsep}
\newcommand{\cdashlinelr}[1]{%
  \noalign{\vskip\aboverulesep
           \global\let\@dashdrawstore\adl@draw
           \global\let\adl@draw\adl@drawiv}
  \cdashline{#1}
  \noalign{\global\let\adl@draw\@dashdrawstore
           \vskip\belowrulesep}}
\makeatother

\usepackage{lipsum}
\usepackage[switch]{lineno}
\usepackage[colorlinks=true,linkcolor=red]{hyperref}

\usepackage[labelsep=period,labelfont=bf]{caption}
\usepackage[nameinlink]{cleveref}  
\crefname{figure}{Fig.}{\textbf{Figure.}}
\crefname{equation}{Eq.}{\textbf{Eq.}}
\crefname{table}{Table}{\textbf{Table.}}
\crefname{section}{Section}{\textbf{Section}}

\newcommand{\lvl}{~~~~}

\begin{document}

\title{StainFuser: Controlling Diffusion for Faster Neural Style Transfer in Multi-Gigapixel Histology Images}

\author{~Robert~Jewsbury$^{1,\ddagger,*}$, ~Ruoyu~Wang$^{1,\ddagger}$, ~Abhir~Bhalerao$^1$,
Nasir~Rajpoot$^{1,2,*}$and~Quoc~Dang~Vu$^{2}$\\$^1$ \{rob.jewsbury, ruoyu.wang.2, abhir.bhalerao, n.m.rajpoot\}@warwick.ac.uk \\ $^2$ qd.vu@histofy.ai \\ $^*$ Corresponding author \\ $\ddagger$ Joint First Authors
\thanks{R.Jewsbury, R.Wang, A.Bhalerao and N.Rajpoot are from the Tissue Image Analytics Centre, Department of Computer Science, University of Warwick, UK}
\thanks{N.Rajpoot and Q.D.Vu are with Histofy Ltd}
}

\maketitle

\begin{abstract}
Stain normalization algorithms aim to transform the color and intensity characteristics of a source multi-gigapixel histology image to match those of a target image, mitigating inconsistencies in the appearance of stains used to highlight cellular components in the images. We propose a new approach, StainFuser, which treats this problem as a style transfer task using a novel Conditional Latent Diffusion architecture, eliminating the need for handcrafted color components. With this method, we curate SPI-2M the largest stain normalization dataset to date of over 2 million histology images with neural style transfer for high-quality transformations. Trained on this data, StainFuser outperforms current state-of-the-art deep learning and handcrafted methods in terms of the quality of normalized images and in terms of downstream model performance on the CoNIC dataset. 
\end{abstract}


\IEEEpeerreviewmaketitle

\section{Introduction}
In recent years, artificial intelligence (AI) algorithms have excelled in many tasks in the Computational Pathology (CPath) domain, such as tumor detection  \cite{gunesli2023federated, Jewsbury2021AQI}, nuclei instance segmentation and classification \cite{hovernet,graham2021conic,xu2023accurate,bashir2020hydramix} and biomarker prediction \cite{wang2023novel,kather2019deep,bashir2023digital}. However, as noted by \cite{tellez2019quantifying,vu2022nuclear,ciompi2017importance, Jahanifar2023DomainGI}, real-life variations often occur during the data acquisition process of gigapixel histology images stained with Haematoxylin and Eosin. These variations, such as stain variance, scanner difference and tissue preparation, can greatly affect the AI algorithms' performance in prognostic and diagnostic assessment of patients. These alterations also pose great challenges for the decision-making of clinical practitioners \cite{salvi2023impact}. Broadly speaking, these alterations can be considered as parts of the bigger domain shift problem in machine learning. Thus, addressing this problem is important for ensuring more consistent results in CPath algorithms and applications.

To address the color variations that occur due to staining and scanner variations, stain normalization is a common approach. At a high level, the aim is to make the color and intensity of a "source" image similar to another image, often termed the "target". Many CPath-specific, handcrafted, methods \cite{ruifrok2001,macenko2009method,vahadane2016structure} have been proposed to separate and recombine the properties and intensities of stains based on their pre-defined chemical properties for capturing light, represented as stain matrices, to align the source image's colors with a desired target image's colors. A stain matrix thusly denotes densities of the stain chemicals within a tissue sample and their corresponding \textit{RGB} values captured in the digital images.

GAN-based methods \cite{cong2022colour,cho2017neural,salehi2020pix2pix} have also been proposed to eliminate the need for these stain matrices. However, training GAN models can be difficult \cite{thanh2020catastrophic, cho2017neural} and thus easily lead to poor generation quality. Additionally, there exists little pairwise image data for training GAN models for stain normalization, many proposed algorithms \cite{cong2022colour,cho2017neural,salehi2020pix2pix} therefore resolve this by \textit{training} their GAN models to reconstruct a \textit{RGB} image from its grayscale counterpart. This approach results in models that are not directly transferable to different domains that exhibit stain properties unseen during training.


We approach the stain normalization problem as a style transfer task introducing a Conditional Latent Diffusion-based architecture for Stain Normalization, termed StainFuser. Recently, diffusion models have emerged as a superior method compared to GANs in both quality and training stability\cite{rombach2022ldm,ho2020denoising,diffusionbeatsgan, ControlNet}. To the best of our knowledge, this is the first study to employ diffusion models for stain normalization which can learn a multi-domain mapping. To train StainFuser, we employ neural style transfer (NST) \cite{neural_style_transfer} to generate the transformed versions of each source and target image pair. This process generates a high-quality dataset and overcomes the paucity of data issue. Thus, we list our contributions as follows:

\begin{itemize}
    \item We propose StainFuser, a novel method that does not require any handcrafted color components (i.e. stain properties) or other transformations and directly applies the style of the target image to the source image.
    \item We publish SPI-2M (Stylized Pathological Images), the largest dataset for stain normalization to date of over 2 million images\footnote{Both our code and data are available at: \url{https://github.com/R-J96/stainFuser}}. We believe this will benefit other generative approaches for stain normalization other than StainFuser.
    \item We demonstrate StainFuser achieves improved image quality compared to the existing state-of-the-art diffusion based model \cite{stainDiff} as well as handcrafted \cite{ruifrok2001, vahadane2016structure} and GAN-based \cite{cong2022colour} methods. Additionally, StainFuser also improves downstream model performance compared to these methods in the CoNIC test set.
    \item We conduct extensive ablation experiments to investigate the importance of components in our model both in terms of image quality, downstream performance and inference time. 
    \item We demonstrate StainFuser's quality on multi-gigapixel Whole Slide Images (WSIs), maintaining consistently high quality across tiles within a WSI.
\end{itemize}

\section{Related Work}
Reinhard \etal \cite{reinhard2001color} introduced a technique for aligning the color distribution of a given image to a reference image in \textit{L*a*b*} color space, which has found applications for stain normalization tasks. However, Reinhard \etal \cite{reinhard2001color} was originally designed for generic color adjustment and was not specifically tailored for histology stain normalization. Subsequently, several prominent approaches in computational pathology, such as Ruifrok \etal \cite{ruifrok2001}, Macenko \etal \cite{macenko2009method} and Vahanade \etal \cite{vahadane2016structure} either proposed or leveraged the concept of the stain matrix to address the task of stain normalization for this research field.

In recent years, GAN-based approaches have emerged as alternatives to the aforementioned handcrafted methods, well-known methods include the works of Salehi \etal \cite{salehi2020pix2pix} and Cong \etal \cite{cong2022colour}. These works follow a vein established by Cho \etal \cite{cho2017neural}. In particular, due to the lack of pairwise data in stain normalization tasks, Cho \etal \cite{cho2017neural} trained their GAN models to reconstruct a \textit{RGB} image from its grayscale counterpart. This grayscale transformation effectively merges diverse stains (or color styles) into a uniform color space \cite{cho2017neural}, and could result in information loss despite employing additional operations \cite{cong2022colour}. Consequently, these models require retraining to adapt to any new target domain with new color distributions. In addition, GAN-based stain normalization models also face challenges in training, notably due to well-known issues such as mode collapse \cite{thanh2020catastrophic} and may need additional constraints for a stabilized generation quality \cite{cho2017neural}.

Recently, denoising diffusion probabilistic models (DDPMs) \cite{Nonequilibrium, ho2020denoising} have emerged as a new set of generative models for image synthesis. DDPMs are a collection of generative models that produce high-quality images through iterative denoising. In contrast to GANs, diffusion models exhibit more stable training and produce higher-quality images \cite{ho2020denoising}. Furthermore, Rombach \etal \cite{rombach2022ldm} enhanced diffusion models' speed and performance by introducing a latent diffusion model (LDM) that operates in variational autoencoder (VAE)-encoded latent space. In addition, the ability to incorporate various conditions (\eg, texts, images, feature representations) into the diffusion models facilitates more applications such as text-to-image generation \cite{imagen, DreamBooth, rombach2022ldm, ControlNet}, image super-resolution \cite{diffusion_supres} or image editing \cite{diffusion_inpainting}. However, the effectiveness of diffusion models in CPath tasks remains under-explored, with limited studies conducted \cite{yellapragada2024pathldm, linmans2024diffusion}. StainDiff \cite{stainDiff} is to our knowledge the first work to use LDMs for stain normalisation; however, it still follows the paradigm of GAN-based approaches where a single domain-to-domain mapping is learnt and requires retraining when used with a new target domain.

\begin{figure*}[t]
\centering
\includegraphics[width=0.9\textwidth]{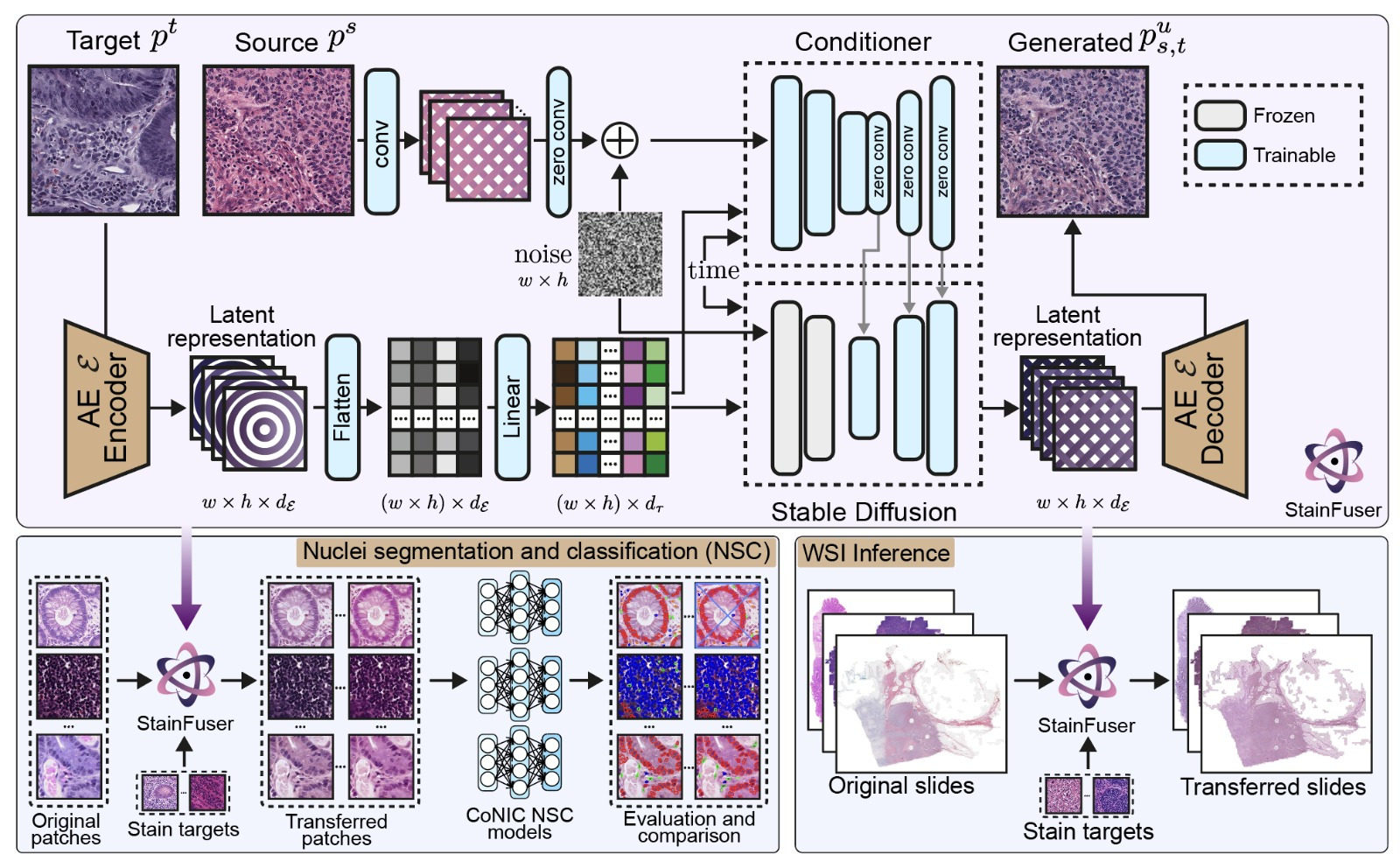}
\caption{The diagram of the proposed StainFuser. StainFuser takes in a source and target image to predict the stain normalized version of the source image. The application of StainFuser was demonstrated through a nuclei segmentation and classification task and a WSI-level inference task.
}
\label{fig:stainfuser}
\end{figure*}

Furthermore, despite numerous new stain normalization methods, their effectiveness on the domain shift problem remains unassessed on a large scale. To the best of our knowledge, the work by Vu \etal \cite{vu2022nuclear}, is the first major attempt to characterize the benefits of stain normalization to a downstream task across a diverse range of stain targets. Specifically, this includes $\sim200$ targets distributed across the color space that typically envelopes CPath image data. Here, the authors compared the performance of Ruifork \cite{ruifrok2001} and Vahadane \cite{vahadane2016structure} methods against the NST method \cite{neural_style_transfer}. They found NST provides the most consistent performance improvement for the nuclei instance segmentation and classification problem, a well-known difficult problem in CPath field \cite{graham2021conic}, across all stain targets.

Thus, inspired by this observation, our paper explores the application of NST to generate pairwise images for training a generative model for stain normalization and explores the utilization of diffusion models for efficient and high-quality stain normalization.

\section{Methodology}
\label{sec:method}
StainFuser aims to predict a Neural Style Transferred version of an input source image given a target image as shown in \Cref{fig:stainfuser}. As no public datasets of sufficient quality and quantity are available, in this section we describe how we curate SPI-2M a pairwise stain normalization dataset from publicly available sources by applying NST to the sampled source and target pairs. Then we detail the architecture and design of StainFuser.

\subsection{Creating SPI-2M}

\label{sec:method:data_curation}
Here, we describe how we curate three distinct image patch sets: the source set $\mathbb{S}=\{p^s_1, p^s_2,...,p^s_n\}$ contains samples to be processed for stain normalization; the target set $\mathbb{T}=\{p^t_1, p^t_2,...,p^t_n\}$ where each sample ideally represents a unique stain variation from the real-world stain distribution; lastly, the transferred set $\mathbb{U}$ is created by applying NST on image pairs from $\mathbb{S}$ and $\mathbb{T}$. 

\subsubsection{Slide Selection} To comprehensively capture and include the real-world variations present in CPath, we retrieved slides from the public TCGA repository\footnote{\url{https://www.cancer.gov/tcga}}. Since the CoNIC challenge dataset \cite{graham2021conic} was used in evaluation, to ensure the consistency of the tissue domain between the training and the evaluation datasets, 3 TCGA cohorts related to the GI tract were selected for our analysis, namely TCGA-STAD (stomach), TCGA-COAD (colon) and TCGA-READ (rectal). Slides and centres used in the CoNIC challenge were excluded. To curate high-quality samples, slides that lack magnification level information and slides scanned at less than 40$\times$ magnification level were also excluded. This results in a total of \num{686} slides scanned at 40$\times$ magnification level for further analysis.

\subsubsection{Patch selection.} Tissue masks of the selected slides were generated using the TIAToolbox \cite{pocock2022tiatoolbox} to remove background, artifacts and pen marks. Subsequently, patches with the size of $1024^2$ at 40$\times$ magnification level were extracted from slides. We denote the patches extracted from these slides as dataset $\mathbb{A}$.

\begin{figure*}[t!]
\centering
\includegraphics[width=\textwidth]{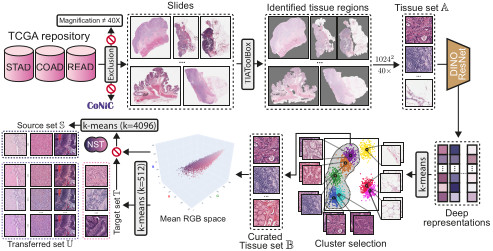}
\caption{
Overview of the data curation workflow: Slides were sourced from the TCGA repository, followed by the patch extraction from identified tissue regions. A two-stage clustering pipeline was implemented to select biologically meaningful and representative patches, ensuring an accurate representation of the real-world morphology and color distribution.
}
\label{fig:data_curation}
\end{figure*}

\subsubsection{Source and target selection.} 
To select representative patches that broadly reflect the diversity of tissue morphology and stains within image set $\mathbb{A}$, we implement a two-stage clustering pipeline, as shown in \Cref{fig:data_curation}. Inspired by \cite{vu2023handcrafted, quiros2023mapping}, we extract biologically meaningful clusters by clustering the deep features of the image patches within $\mathbb{A}$.

In the first stage, using ResNet-50 pretrained with DINO \cite{caron2021emerging} on the ImageNet dataset \cite{2015imagenet}, for each patch $p \in \mathbb{A} $, we obtain a set of deep feature vectors $Z=\{z_1, z_2,...,z_n\}$ from the images in tissue set $\mathbb{A}$. We then use k-means clustering to retrieve a set $C=\{c_1, c_2,...,c_{128}\}$ of 128 clusters from the feature set $Z$. Afterward, we visually examine the patches within each cluster to determine if that cluster contains unfit tissue components such as adipose tissues or more meaningful histological patterns (\ie majorly containing known tissue patterns like glands or lymphoid aggregate). Subsequently, we remove all patches within the cluster we deem unfit from further consideration and denote the set containing the remaining valid image patches as set $\mathbb{B}$.

In the second stage, to select representative patches that reflect the staining style (\ie the color), instead of using deep features to represent each patch as in the first stage, we represent each patch within $\mathbb{B}$ by their mean RGB value $\hat{z}$ and obtain $\hat{Z}=\{\hat{z}_1, \hat{z}_2,...,\hat{z}_n\}$. 


For curating the target set $\mathbb{T}$, we perform k-means clustering on $\hat{Z}$ and obtain a set of 512 clusters $\hat{C}=\{\hat{c}_1, \hat{c}_2,...,\hat{c}_{512}\}$. To select the most representative patch of each cluster, we select \textit{one single patch} within $\mathbb{B}$ which is the closest to its cluster center in $\hat{C}$ in terms of the Euclidean distance in the RGB color space.

On the other hand, for curating the source set $\mathbb{S}$, we first obtain a subset $\bar{\mathbb{B}}=\{p \in \mathbb{B} : p \notin \mathbb{T}\}$ before performing the same clustering and the patch selection. Here, we extract 4096 clusters and similarly select \textit{one single patch} within $\bar{\mathbb{B}}$ to represent each cluster.

In summary, from $\mathbb{A}$, we obtained the source tissue set $\mathbb{S}$ which contains \num{4096} images and the target tissue set $\mathbb{T}$ which contains \num{512} images that are evenly spaced in the color space of $\mathbb{A}$ (\ie TCGA-STAD, TCGA-COAD and TCGA-READ).

\subsubsection{Neural Style Transfer}
\label{sec:neural_style_transfer}
To generate the training data we perform NST \cite{neural_style_transfer} with our sampled source set $\mathbb{S}$ and target set $\mathbb{T}$. Specifically, we treat a given source image $p^s \in \mathbb{S}$ as the content image, a given target image $p^t \in \mathbb{T}$ as the style image and generate a stylized image $p^u_{s,t}$. At the start of the NST process, $p^u_{s,t}$ is a clone of the content image \ie $p^u_{p^s,p^t} = p^s_i$ which is then refined by the NST process. Using a VGG16 pre-trained on ImageNet, denoted as $\bar{F}$, we extract features from every pooling layer in the network for all images creating three sets of features $\bar{F}_{s}$, $\bar{F}_{t}$ and $\bar{F}_{u}$ where $\bar{F}_i=\left\{\bar{f}^q, \forall q \in \{1,2,...,n\} \right\}$ where $n$ is the number of pooling layers in the VGG16 and $f^q$ is the feature representation of image $i$ at layer $q$.

Given $\bar{F}_s$, $\bar{F}_t$ and $\bar{F}_u$ we compute the mean squared loss between $\bar{F}_s$ and $\bar{F}_u$ feature-wise at each layer resulting in the overall content loss across all pooling layers
\begin{equation} \label{eq:contentLoss}
    L_{content}(\bar{F}_s, \bar{F}_u) = \sum_0^n(\bar{F}_s-\bar{F}_u)^2.
\end{equation}
The style loss is computed by calculating the Gram matrix $G$ of the target image's features $\bar{F}_t$ and the stylized image $\bar{F}_u$ at each layer and computing the mean squared loss between these Gram matrices
\begin{equation} \label{eq:styleLoss}
    L_{style}(\bar{F}_t, \bar{F}_u) = \sum_0^n(G(\bar{F}_t) - G(\bar{F}_u))^2.
\end{equation}
The final overall loss is given by
\begin{equation} \label{eq:nstLoss}
    L_{total} = \alpha L_{content} + \gamma L_{style},
\end{equation}
where $\alpha$ and $\gamma$ are weighting constants.

We set $\alpha$ to be 1 and $\gamma$ to be \num{10000} for all of our work as this was found to lead to the best qualitative results. This loss is then backpropagated through the stylized image, $p^u$, for 300 iterations producing the final version of $p^u$. For a $1024^2$ RGB image, this works out to be \num{3145728} parameters.

We use the Adam \cite{Kingma2014AdamAM} optimizer and mixed precision to increase the data generation speed due to the significant computational cost of this process. By repeating this process for every pairing of every image in $\mathbb{S}$ and $\mathbb{T}$ we generate the corresponding set $\mathbb{U}$ where every $s_i$ has been transformed with the style of every $t_j$. In total this results in \num{2097152} images for training.

Additionally, we scale the matrix dot product operation in the gram matrix calculation $G$ while using mixed precision to prevent float overflow error that occurs during the transition between fp16 and fp32. Empirically, we found that NST with fp16 provides the same image quality as NST with fp32, with an average cosine similarity of 0.999 across 10 image pairs. Using fp16 instead of 32 provides a speedup of 1.25 to 2 depending on the GPU used for NST.

\subsubsection{Generating the transferred set.}
Finally, we apply NST on each pairwise combination of image patches in $\mathbb{S}$ and $\mathbb{T}$. Through this process, for a given pair $p^s, p^t$ we obtain image $p^u_{s,t}=NST(p^s, p^t)$ whose tissue components are the same as $p^s$ but have their color based on the stain of similar tissue morphology observed in $p^t$. This process results in \num{2097152} style transferred images in the transferred set $\mathbb{U}$.


\subsection{Design of StainFuser}
\subsubsection{Latent Diffusion Models}
\label{sec:latent_diffusion_models}
Latent diffusion models (LDMs) \cite{rombach2022ldm}, like other DDPMs, consists of a forward and a reverse process. However, a distinctive feature of LDM is that it operates in the latent space, encoded via an AutoEncoder $\mathcal{E}$, instead of in the pixel space. This significantly improves the efficiency of the diffusion process. Therefore, LDM has been adopted in this study. The forward diffusion process of LDM is defined as a Markov chain which maps the sample from the real data distribution to a Gaussian distribution by gradually adding Gaussian noise to the sample. Let $z_0$ denote the encoded latent representation of the input image $p$, obtained by an AutoEncoder $\mathcal{E}$, such that $z_0= \mathcal{E}(p)$; while $z_t$ denote the noised version of $z_0$ at timestep $t$, the forward diffusion process $q(\cdot)$ is defined as 
\begin{equation} \label{eq:forward}
q(z_t|z_{t-1}):=\mathcal{N}(z_t; z_{t-1}\sqrt{1-\beta_t}, \beta_tI),
\end{equation}
where $\{\beta_t \in (0,1)\}^T_{t=1}$ is the time scheduler and $I$ is the identity matrix. The time scheduler $\beta_t$ controls the amount of noise to be added to the sample $z_{t-1}$ at timestep $t$. The reverse process aims to reconstruct the initial latent representation $z_0$ from $z_T$. This is achieved by training a time conditional model to estimate the conditional probability distribution to recover the latent representation $z_{t-1}$ at timestep $t-1$ given $z_t$. In LDM, a time-conditional UNet \cite{unet} is used as the backbone network for such purpose. If $\beta_t$ is small enough, $q(z_{t-1}|z_t)$ will also be a Gaussian distribution \cite{Nonequilibrium}. Therefore, the reverse process can be defined as
\begin{equation} \label{eq:reverse}
p_\theta(z_{t-1}|z_t):=\mathcal{N}(z_{t-1}; \mu_\theta(z_t,t), \Sigma_\theta(z_t,t)),
\end{equation}
where $\mu_\theta(z_t,t)$ and $\Sigma_\theta(z_t,t)$ is the mean and the covariance of the Gaussian distribution determined by time $t$, latent $z_t$ at timepoint $t$, and the learned model parameters $\theta$.

\subsubsection{StainFuser Architecture}
StainFuser adapts a pre-trained Stable Diffusion (SD) Latent Diffusion Model (LDM) for neural style transfer (NST) in histopathology images, \Cref{fig:stainfuser}. The model takes a source image patch $p^s$ and a target-stain image patch $p^t$ as inputs, generating a transferred sample $p^u_{s,t}$ that retains the structure of $p^s$ while applying the stain characteristics of $p^t$.

Modifications to the SD model include:
\begin{enumerate}
    \item Input Adaptation: The text-encoder part of the CLIP encoder is replaced with an additional VAE embedding $\mathcal{E}(\cdot)$ to accept image input $p^t$.
    \item Embedding Processing: The embedded target image $\mathcal{E}(p^t)\in \mathbb{R}^{w \times h \times d_\mathcal{E}}$ is flattened to $\mathcal{E}^{\prime}(p^t)\in \mathbb{R}^{(w \times h) \times d_\mathcal{E}}$ and projected through a linear layer $l(\cdot)$ to $l(\mathcal{E}^{\prime}(p^t)) \in \mathbb{R}^{(w \times h) \times d_\tau}$, ensuring compatibility with the SD U-Net architecture.
    \item Cross-Attention Integration: The projected representation is incorporated into the UNet backbone using cross-attention layers \cite{rombach2022ldm}:
    \begin{equation} \label{eq:attention}
    \text{Attention}(Q, K, V) := \text{softmax}\left(\frac{QK^T}{\sqrt{d}}\right) \cdot V,
    \end{equation}
    where $Q=W_Q^{(i)}\cdot \varphi_i(z_t)$, $K=W_V^{(i)}\cdot l(\mathcal{E}^{\prime}(p^t))$, and $V=W_V^{(i)}\cdot l(\mathcal{E}^{\prime}(p^t))$. Here, $\varphi_i(z_t)$ denotes an intermediate output of the UNet, and $z_t$ denotes the noised version of $z_0=\mathcal{E}(p^u_{s,t})$ at timestep $t$.
    \item Source Image Control: To maintain the structure of $p^s$, it is incorporated using zero convolution layers following Zhang \etal's approach\cite{ControlNet} into a trainable copy of the original SD ($\mathcal{F}(\cdot;\Theta_c)$). This involves:
    \begin{itemize}
        \item Encoding $p^s$ with a learnable network $h$.
        \item Creating trainable copies of SD blocks, $\mathcal{F}(\cdot;\Theta_c)$.
        \item Incorporating the encoded source image through these zero convolution layers.
    \end{itemize}
    \item Conditioner: Processes the timestep $t$, encoded target-stain image, and concatenated source image with noise vector to generate intermediate representations for the SD model.
\end{enumerate}

The learning objective of StainFuser is defined as:
\begin{equation} \label{eq:loss}
\mathcal{L} = \mathbb{E}{z^u_0, t, p^s, p^t, \epsilon \sim \mathcal{N}(0,1)} \Big[\lVert \epsilon - \epsilon\theta (z^u_t, t, p^s, p^t) \lVert_2^2 \Big],
\end{equation}
where $z^u_0$ and $z^u_t$ are latent representations of $p^u_{s,t}$ at timesteps 0 and $t$, $\epsilon$ is the input noise, and $\epsilon_\theta (\cdot)$ is the estimated noise by the diffusion model. We freeze the encoder part of the original SD backbone but train all other components of the overall StainFuser architecture.

\begin{figure*}[t]
\centering
\includegraphics[width=0.90\textwidth]{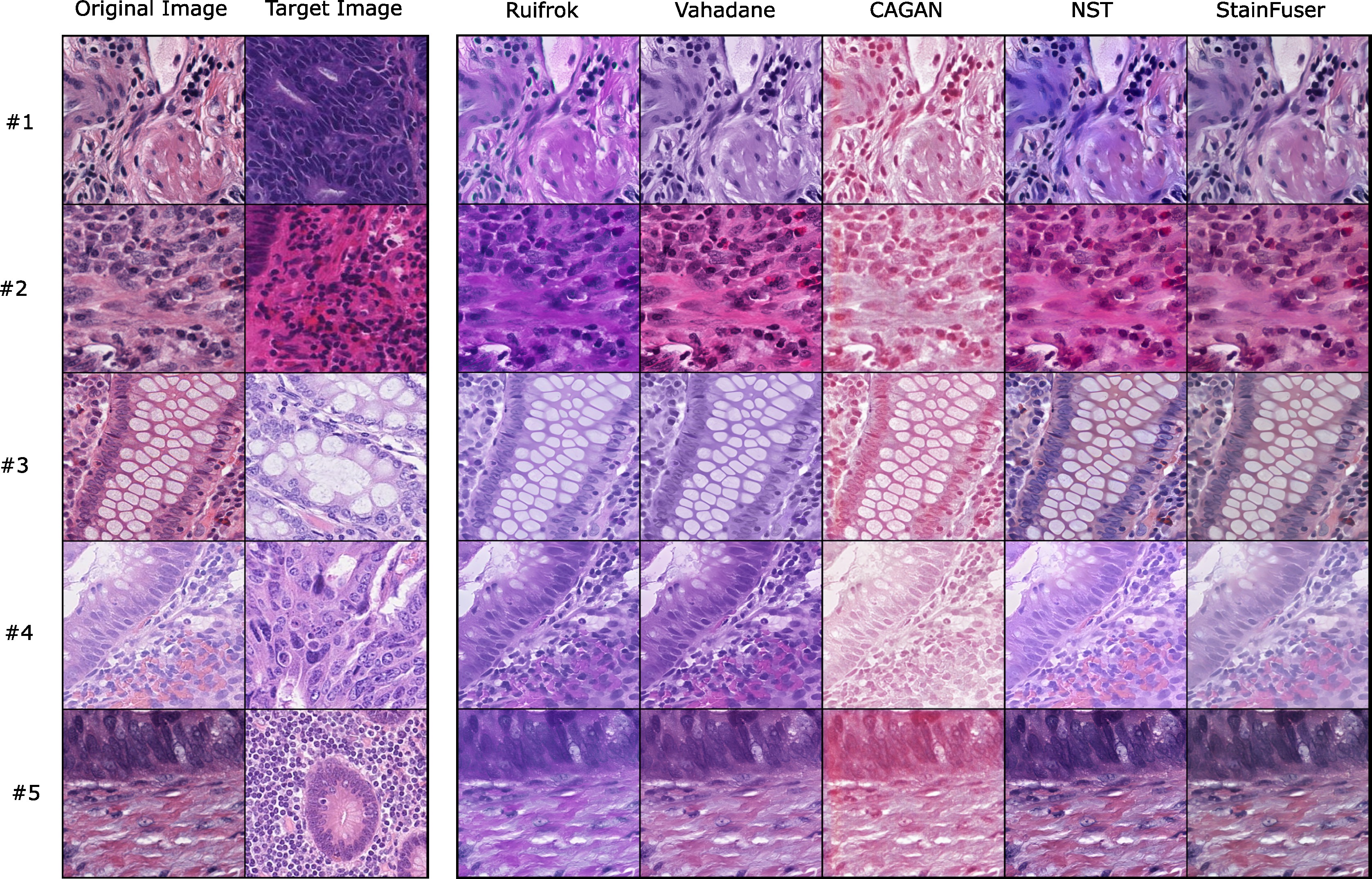}
\caption{Qualitative comparisons between StainFuser and other methods on CoNIC test set examples. All inference was performed at $512^2$ resolution and then resized for display purposes. Only StainFuser and NST preserve the color contrast between important tissue components such as stroma, glands, lumen and blood vessels present in the original image.
}
\label{fig:qualOverallMicro}
\end{figure*}

\section{Experiments}
\label{sec:experiments}

In our experiments, we compare StainFuser with two traditional stain normalization methods (Ruifrok \cite{ruifrok2001} and Vahadane \cite{vahadane2016structure}), a GAN-based method (CAGAN \cite{cong2022colour}) and NST \cite{neural_style_transfer} itself in terms of image quality and downstream performance for nuclei instance segmentation and classification on the CoNIC dataset \cite{Graham2023CoNICCP}. We also compare with the first LDM based stain normalization model, StainDiff's \cite{stainDiff} published results in terms of image quality. We perform extensive ablations of the training and inference hyperparameters including qualitative results. Finally, we also present results applying the methods for WSI inference showcasing the clinical applications of StainFuser and detail the limitations of our approach. Training details such as hyperparameters and other observations can be found in the \Cref{sec:suppImpl}.

\subsection{Evalution Datasets}
We trained our StainFuser models based on the curated dataset as described in \Cref{sec:method:data_curation}. To evaluate our models, we primarily utilized the data from the CoNIC challenge \cite{Graham2023CoNICCP}. This dataset consists of H\&E stained image tiles from colorectal cancer WSIs, there are 5k training images with \num{431913} unique nuclei instances and 1k testing images with \num{103150} nuclei instances. Each image is annotated with panoptic segmentation labels of 6 nuclei classes, neutrophils, epithelial cells, lymphocytes, plasma cells, eosinophils and connective cells in addition to the background. As noted in \Cref{sec:method:data_curation}, the curated data for training our proposed StainFuser does not include any examples within the CoNIC data. We also used the MITOS-ATYPIA 14 dataset \footnote{\url{https://mitos-atypia-14.grand-challenge.org}} sourced from breast tissue scanned with two different scanners, Aperio Scanscope XT and Hamamatsu Nanozoomer 2.0-HT, to compare with other approaches in terms of image quality. We follow the same experimental settings as prior work \cite{stainDiff} and randomly crop 500 paired patches from slides in the test set.

\begin{figure*}[t]
\centering
\includegraphics[width=0.90\textwidth]{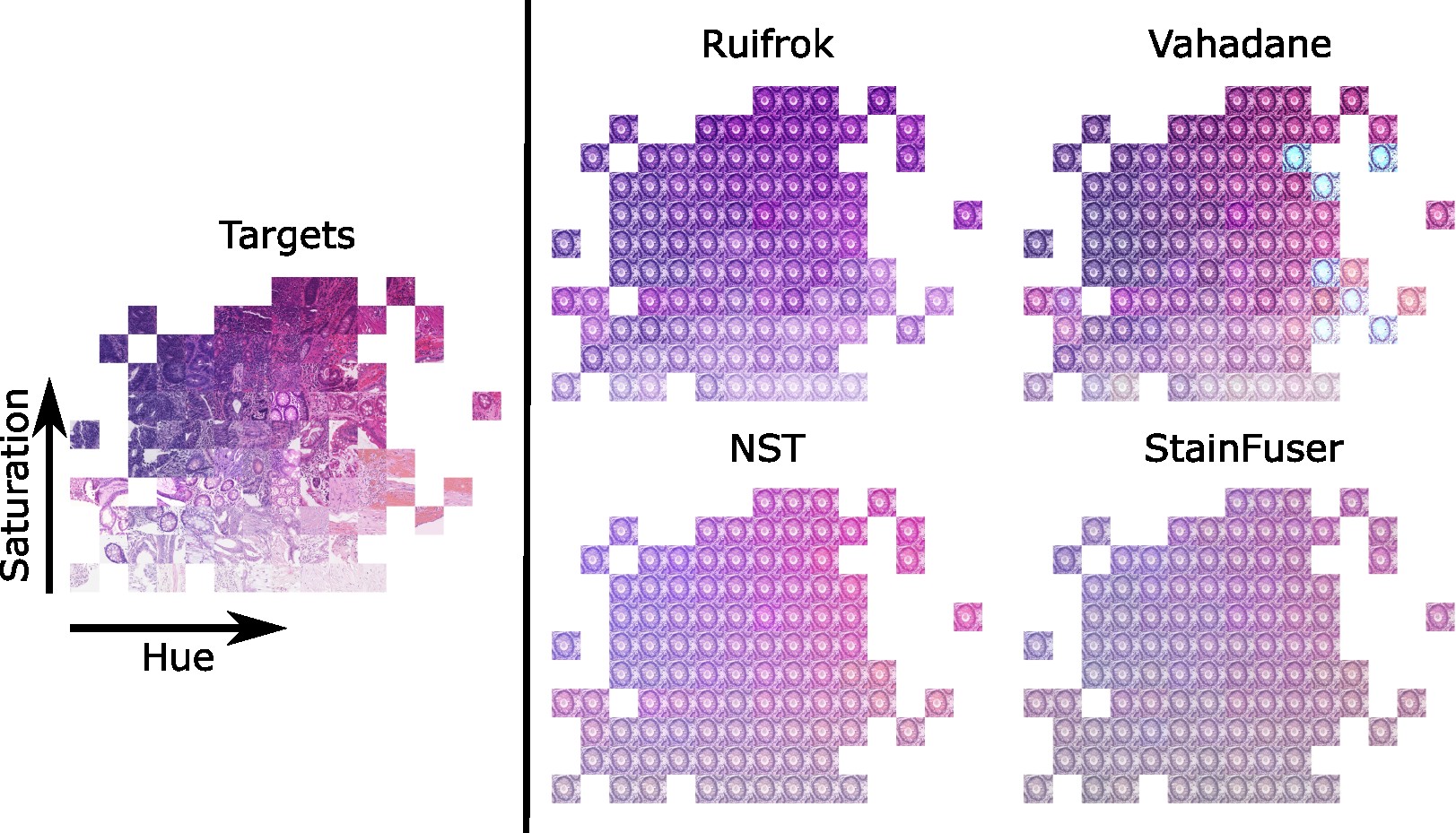}
\caption{Target images selected by sampling in HSV space and a test sample normalised by each method assessed. Targets are displayed on 2D plane where x-axis is Hue and y-axis is Saturation by the mean value of the respective target's Hue and Saturation. High-resolution versions of each set of images are included in the \Cref{sec:addResults}.
}
\label{fig:qualOverallMacro}
\end{figure*}

\subsection{Experimental Settings}
Inspired by Vu \etal \cite{vu2022nuclear}, we followed the same setup for evaluation on CoNIC for both image quality and downstream analysis. Specifically, we up-scaled the test data from CoNIC \cite{Graham2023CoNICCP} with ESRGAN \cite{esrgan} super-resolution creating the \textbf{Control} set. These images were $\num{1024}^2$ and used for NST as this has been shown to significantly improve the performance of NST \cite{vu2022nuclear, neural_style_transfer}. This \textbf{Control} set was used for all comparisons with the original data and resized back to $256^2$ or $512^2$ with bi-linear interpolation to make the comparisons between methods as fair as possible. For each method and experimental setting studied we normalized the entirety of the \textbf{Control} version of the test set \textit{w.r.t.} each sampled target. \ie for each sampled target we generate a new version of each image in the testing set where the given image has been normalized using the chosen method with respect to the specified target.

This process is designed to provide a robust evaluation of the stain normalization process, instead of assessing stain normalization methods for one target image only. Existing work has shown that, for example, Vahadane stain normalization can lead to a wide spread of performance downstream depending on what image target is chosen \cite{vu2022nuclear}. While the exact mechanism by which this variation arises has not been fully explored we believe the principle of assessing performance across a range of sampled targets provides a more thorough and representative evaluation of downstream performance compared to using one single target which can be cherry-picked easily.

\subsubsection{Motivation}
We further argue that assessing normalization methods on image-level tasks such as tumor or tissue classification is insufficient to fully assess the important capabilities of a stain normalization algorithm. Most of the publicly available datasets used for these tasks such as Kather100k \cite{kather2019deep} and BreakHis \cite{Spanhol2016BreakHis} are image-level classification tasks where each image is assigned one of $n$ labels representing the class of the image such as tumor vs. non-tumor classification. It is relatively easy to achieve high performance on these image-level classification tasks using features that do not take account of local morphology, mean color for example. This implies that when assessing with a framework such as this a normalization method could theoretically disrupt the local morphology of the original image and still achieve superior performance compared to a baseline because it aligns the unseen sample better to the features (abstract or not) a model has learned. This would be potentially disastrous in clinical applications where the local morphology of nuclei is highly significant for many clinical tasks. Instance-level nuclei segmentation tasks like CoNIC, however, do not suffer from this issue. If a normalization method perturbs the original morphology during the style transfer process either a downstream model will not be able to detect a given nucleus, missing the instance, or it will segment the distorted nucleus resulting in a contour boundary with poor intersection compared to the ground truth mask and thus panoptic segmentation metrics will penalize this accordingly, provided the downstream model has good performance. Prior work \cite{vu2022nuclear} has shown modern nuclei instance segmentation and classification models are robust regarding compression artifacts; as such we argue they are suitable for this task.

\subsubsection{Comparisons}
We compare the StainFuser normalised versions of the same 101 versions of the CoNIC test set used by Vu \etal \cite{vu2022nuclear} against the reported results of Ruifrok \cite{ruifrok2001}, Vahadane \cite{vahadane2016structure} and NST \cite{neural_style_transfer}. Additionally, we also compared against the published CAGAN method \cite{cong2022colour} trained on TCGA-IDH \cite{cagan_tcgaidh} as the authors claimed CAGAN could normalize images from other datasets and sources. We evaluate StainFuser and the other stain normalisation algorithms in terms of image quality and also explore them for downstream use. We utilised the nuclei instance segmentation and classification task in CoNIC \cite{Graham2023CoNICCP} for this and assessed the performance of three state-of-the-art (SoTA) methods from the CoNIC challenge, namely \textit{Pathology AI} (PathAI) \cite{pathologyai2022conic}, \textit{MDC Berlin $\vert$ IFP Bern} (Bern) \cite{mdc2022conic} and \textit{EPFL $\vert$ StarDist} (StarDist) \cite{stardist2022conic} on each stain normalized test set individually. This was to evaluate and compare StainFuser across model architectures as well as target images for this challenging downstream task. For evaluating the model performance, we utilized $m\mathcal{PQ^+} AUC$ as described in \cite{vu2022nuclear}. Where possible, we also report the mean $m\mathcal{PQ^+} AUC$ $\pm$ the standard deviation of a model's downstream performance across the entire distribution of altered test sets.

\subsection{Evaluating Image Appearance - Qualitative} 
\label{sec:exp:qual}

At the micro level, \Cref{fig:qualOverallMicro} shows that Ruifrok \cite{ruifrok2001} produces very purple images regardless of the chosen target. On the other hand, while Vahadane does not suffer from this issue, it fails to differentiate the color of distinct cellular components. For instance, in \#3, the inner portion of the gland is also colored purple. The worst of all is CAGAN \cite{cong2022colour} where it can not utilize the target images and can only map to a single domain (rose red). Unlike these, StainFuser and NST produce images that maintain good contrast between important cellular components, such as the stroma and lumen as in \#3. Compared to NST, visually, StainFuser produces more color-consistent images but its colors are less vibrant, as seen in \#1.

At the macro level, we display sampled target images in \Cref{fig:qualOverallMacro}. By normalizing a single sample image using these chosen targets, we can evaluate how each normalization method performs across a typical colorspace of CPath data. From \Cref{fig:qualOverallMacro}, we see that Vahadane has many irregular outputs, such as the orange outputs and very pale images produced in the bottom row. Ruifrok is consistent in terms of output color, which is predominantly purple; however, it struggles when very pale, light images are used as the target (bottom region of the plot). NST and StainFuser however, produce more consistent normalized images across the evaluated color range. Compared to StainFuser, NST produces more vibrant images in general.

Overall, our results in \Cref{fig:qualOverallMicro} and \Cref{fig:qualOverallMacro} demonstrate that StainFuser has comparable performance against NST and is superior to other methods. It also is capable of producing images that are color-consistent with a highly varied range of target images, unlike handcrafted methods such as Ruifrok and Vahadane normalization.

\subsection{Evaluating Image Appearance - Quantitative}

\begin{table*}[t!]
\caption{Image quality comparisons with the SoTA methods on the CoNIC test set. All results are reported for $512^2$ images. Inference time is reported per image, Ruifrok and Vahadane times were computed on an Intel Xeon Gold 6240 CPU multiprocessing 32 images simultaneously; NST, CAGAN and StainFuser were all computed on an A100 GPU with a batch size of 32 for CAGAN and StainFuser. All times were calculated over a full test set of 1000 images, the time per batch of 32 images was recorded and then averaged and reported along with the standard deviation across batches. Best results are shown in {\color[HTML]{0000FF}{blue}}.}
\label{tab:img-quality}
\centering
\small
\resizebox{0.75\textwidth}{!}{
\begin{NiceTabular}{ccccc}[]
\toprule
\RowStyle{\bfseries}
\Block[c]{}{Method} & \Block[c]{}{Inference time (s)} & \Block[c]{}{FID ($\downarrow$)}     & \Block[c]{}{PSNR ($\uparrow$)}    & \Block[c]{}{SSIM ($\uparrow$)} \\
\midrule
Ruifrok \cite{ruifrok2001} &  \phantom{1}0.215 $\pm$ 0.017 & \lvl 34.261 $\pm$ \phantom{1}4.848 & \lvl 14.395 $\pm$ 1.298 & \lvl 0.855 $\pm$ 0.039 \\
Vahadane \cite{vahadane2016structure} & \phantom{1}0.518 $\pm$ 0.051 & \lvl 37.010 $\pm$ 18.393 & \lvl 14.363 $\pm$ 1.435 & \lvl 0.844 $\pm$ 0.063 \\
CAGAN \cite{cong2022colour} & {\color[HTML]{0000FF}{\phantom{1}0.021 $\pm$ 0.006}} & \lvl 119.789 & \lvl 16.653 & \lvl 0.847 \\
NST \cite{neural_style_transfer} &  12.404 $\pm$ 1.184 & \lvl {\color[HTML]{0000FF}{22.210 $\pm$ \phantom{1}8.561}} & \lvl {\color[HTML]{0000FF}{24.937 $\pm$ 3.202}} & \lvl {\color[HTML]{0000FF}{0.931 $\pm$ 0.020}} \\
\cdashlinelr{1-5}
\textbf{StainFuser} & \phantom{1}0.413 $\pm$ 0.005  & \lvl 25.882 $\pm$ \phantom{1}8.233 & \lvl 23.911 $\pm$ 0.816 & \lvl 0.875 $\pm$ 0.010 \\
\bottomrule
\end{NiceTabular}
}
\end{table*}

\subsubsection{CoNIC Comparisons}
Following traditional approaches for evaluating generative models, we compute the Fr\'echet Inception Distance (FID) \cite{FID}, Peak Signal to Noise Ratio (PSNR) and Structural Similarity Index Measure (SSIM) \cite{SSIM} for the generated test set(s). Our results are detailed in \Cref{tab:img-quality} along with inference time comparisons. We find StainFuser outperforms Ruifrok, Vahadane and CAGAN in terms of FID, PSNR and SSIM. While NST has superior image quality, StainFuser is competitive and substantially faster achieving a 30$\times$ speed up in inference time. 

\begin{table*}
\caption{Image quality comparisons on Atypia-14. Best results are shown in {\color[HTML]{0000FF}{blue}}.}
\label{tab:stainDiffComp}
\centering
\small
\resizebox{0.75\textwidth}{!}{
\begin{NiceTabular}{cccc}[]
\toprule
\RowStyle{\bfseries}
\Block[c]{}{Method} & \lvl \Block[c]{}{PC ($\uparrow$)} & \Block[c]{}{SSIM ($\uparrow$)} & \Block[c]{}{FSIM ($\uparrow$)}\\
\midrule
Vahadane &\lvl 0.561 $\pm$ 0.058 &\lvl 0.639 $\pm$ 0.063 &\lvl 0.710 $\pm$ 0.031\\
StainDiff~\cite{stainDiff}\tabularnote{Results are taken from the original paper.} &\lvl 0.599 $\pm$ 0.025 &\lvl 0.721 $\pm$ 0.017 &\lvl 0.753 $\pm$ 0.010\\
\cdashlinelr{1-4}
\textbf{StainFuser} &\lvl {\color[HTML]{0000FF}{0.910 $\pm$ 0.019}} &\lvl {\color[HTML]{0000FF}{0.753 $\pm$ 0.029}} &\lvl {\color[HTML]{0000FF}{0.858 $\pm$ 0.017}}\\
\bottomrule
\end{NiceTabular}
}
\end{table*}
\subsubsection{Atypia-14 Comparisons}
We follow prior work and compute the Pearson correlation coefficient (PC), Structural Similarity Index Measure (SSIM) and Feature Similarity Index for Image Quality Assessment (FSIM). We display our results in \Cref{tab:stainDiffComp} finding that StainFuser substantially outperforms other stain normalisation methods. It's worth noting that Atypia-14 is in breast tissue an organ site entirely unseen to StainFuser while the other methods listed are trained directly in that domain or perform pairwise mapping

\subsection{Downstream Evaluations}
We report our downstream results in \Cref{tab:main results}. Similar to our results in \Cref{sec:exp:qual}, we observe that StainFuser consistently outperforms Ruifrok, Vahadane and CAGAN for all models and all metrics. Interestingly, compared to NST (\ie the ground truth for training StainFuser), StainFuser outperforms NST in terms of $m\mathcal{PQ^+} AUC$ when using the Bern model. On the other hand, for both PathAI and StarDist, NST is better than StainFuser in terms of $m\mathcal{DQ^+} AUC$ and $m\mathcal{PQ^+} AUC$. We include qualitative examples of each model's performance with each normalization method in the supplementary material in \Cref{fig:qualPathAI}, \Cref{fig:qualBern} and \Cref{fig:qualStarDist}.

Our results thus show that different model architectures and training strategies respond differently to various normalization methods at inference time. Additionally, the superiority of NST and StainFuser compared to other methods in image quality and consistency also is reflected in the downstream evaluation. This is evidenced by a clear gap in performance across the board between NST and StainFuser and other methods.


\subsubsection{Results per target}
\begin{table*}[t!]
\caption{Comparison with other Stain Normalisation methods as a test-time augmentation on the CoNIC test set. Results are the mean $\pm$ standard deviation across all 101 target sets except for CAGAN as there was no distribution of results for this model. The best-performing stain normalisation method is highlighted in {\color[HTML]{0000FF}{blue}}.}
\label{tab:main results}
\centering
\small
\resizebox{0.75\textwidth}{!}{
\begin{NiceTabular}{ccccc}[]
\toprule
\RowStyle{\bfseries}
\Block[c]{}{Model} & \Block[c]{}{Method} & \Block[c]{}{$m\mathcal{DQ^+} AUC (\uparrow)$} & \Block[c]{}{$m\mathcal{SQ^+} AUC (\uparrow)$}  & \Block[c]{}{$m\mathcal{PQ^+} AUC (\uparrow)$}\\
\midrule
 \Block{5-1}{\textbf{PathAI}}&\lvl Ruifrok \cite{ruifrok2001} & \lvl 0.248 $\pm$ 0.012 &\lvl 0.374 $\pm$ 0.002 &\lvl 0.186 $\pm$ 0.010 \\
 &\lvl Vahadane \cite{vahadane2016structure} &\lvl 0.240 $\pm$ 0.069 &\lvl 0.368 $\pm$ 0.015 &\lvl 0.179 $\pm$ 0.052\\
  &\lvl CAGAN \cite{cong2022colour} & 0.163 & 0.366 & 0.121\\
 &\lvl NST \cite{neural_style_transfer} &\lvl {\color[HTML]{0000FF}{0.287 $\pm$ 0.016}} &\lvl 0.375 $\pm$ 0.001 &\lvl {\color[HTML]{0000FF}{0.215 $\pm$ 0.012}} \\
 \cdashlinelr{2-5}
 &\lvl \textbf{StainFuser} &\lvl 0.283 $\pm$ 0.010 &\lvl {\color[HTML]{0000FF}{0.378 $\pm$ 0.001}} &\lvl 0.211 $\pm$ 0.007\\
\midrule
\Block{5-1}{\textbf{Bern}} &\lvl Ruifrok \cite{ruifrok2001} &\lvl 0.275 $\pm$ 0.010 &\lvl 0.379 $\pm$ 0.003 &\lvl 0.209 $\pm$ 0.009\\
&\lvl Vahadane \cite{vahadane2016structure} &\lvl 0.268 $\pm$ 0.031 &\lvl 0.380 $\pm$ 0.004 &\lvl 0.205 $\pm$ 0.024\\
&\lvl CAGAN \cite{cong2022colour} & 0.187 & 0.379 & 0.143 \\
&\lvl NST \cite{neural_style_transfer} &\lvl 0.294 $\pm$ 0.004 &\lvl 0.382 $\pm$ 0.001 &\lvl 0.225 $\pm$ 0.004 \\
\cdashlinelr{2-5}
&\lvl \textbf{StainFuser} &\lvl {\color[HTML]{0000FF}{0.294 $\pm$ 0.003}} &\lvl {\color[HTML]{0000FF}{0.392 $\pm$ 0.001}} &\lvl {\color[HTML]{0000FF}{0.225 $\pm$ 0.003}} \\
\midrule
\Block{5-1}{\textbf{StarDist}}&\lvl Ruifrok \cite{ruifrok2001} & \lvl 0.271 $\pm$ 0.006 &\lvl 0.382 $\pm$ 0.002 &\lvl 0.208 $\pm$ 0.004\\
&\lvl Vahadane \cite{vahadane2016structure} &\lvl 0.249 $\pm$ 0.054 &\lvl 0.380 $\pm$ 0.004 &\lvl 0.191 $\pm$ 0.041\\
&\lvl CAGAN \cite{cong2022colour} & 0.189 & 0.387 & 0.149 \\
&\lvl NST \cite{neural_style_transfer} &\lvl {\color[HTML]{0000FF}{0.280 $\pm$ 0.008}} &\lvl 0.384 $\pm$ 0.001&\lvl {\color[HTML]{0000FF}{0.216 $\pm$ 0.006}}  \\
\cdashlinelr{2-5}
&\lvl \textbf{StainFuser} &\lvl 0.274 $\pm$ 0.005 &\lvl {\color[HTML]{0000FF}{0.392 $\pm$ 0.001}} &\lvl 0.211 $\pm$ 0.004\\
\bottomrule
\end{NiceTabular}
}
\end{table*}

\begin{figure*}[t]
\centering
\includegraphics[width=0.85\textwidth]{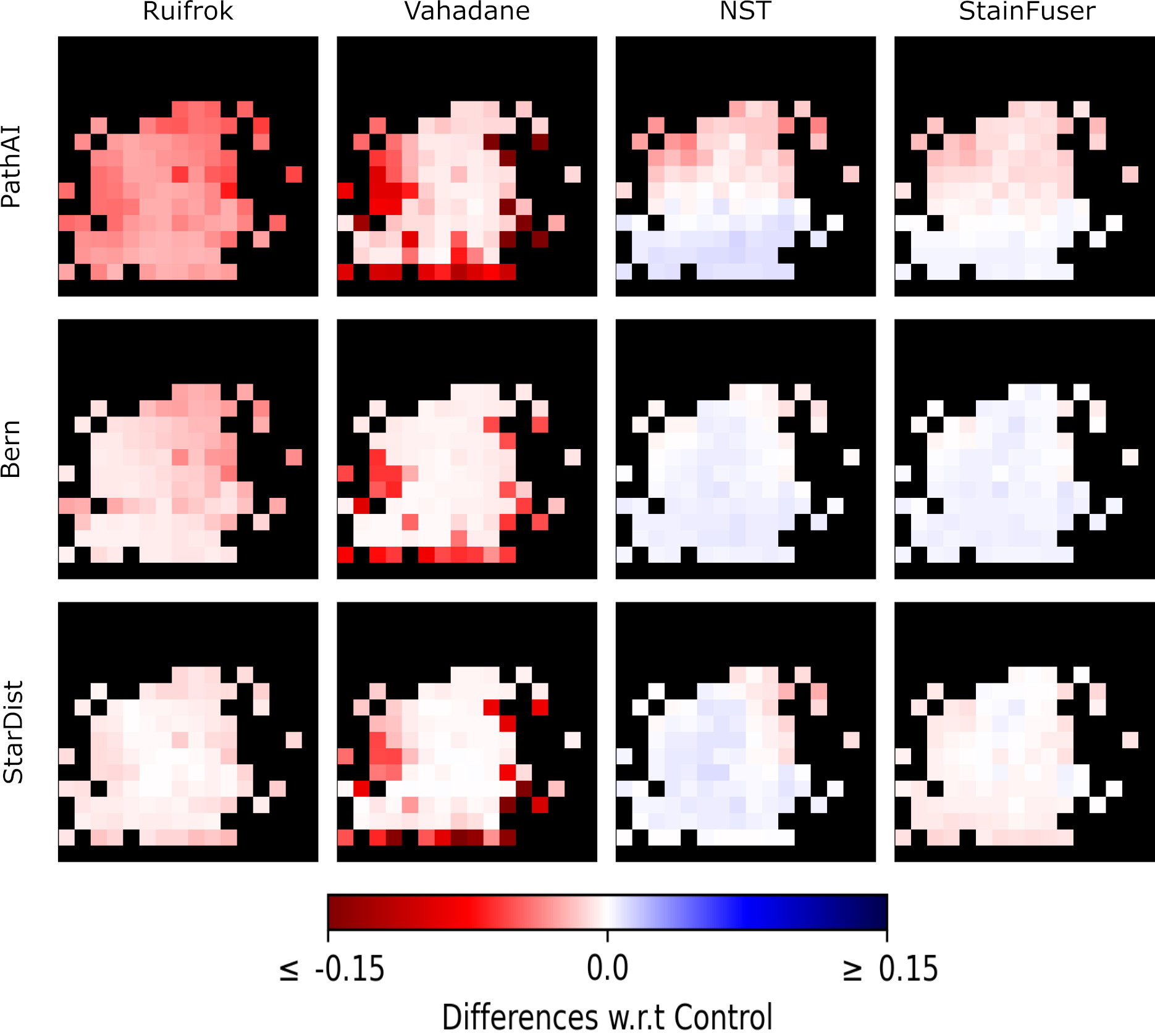}
\caption{Heatmaps of the difference in the $m\mathcal{PQ^+} AUC$ between the \textbf{Control} and the test set where its color was shifted w.r.t each sampled target. Changes in performance are displayed in the same pattern as their corresponding target in \Cref{fig:qualOverallMacro}. CAGAN is excluded as it can not normalize w.r.t. a specific target image.
}
\label{fig:heatmap}
\end{figure*}
To explore how model performance varied by normalization target across methods we display the individual results from \Cref{tab:main results} as a set of heatmaps in \Cref{fig:heatmap}. Each square cell is located at the corresponding position of the target in HSV space from \Cref{fig:qualOverallMacro}. The color of each cell denotes the relative performance in terms of $m\mathcal{PQ^+} AUC$ of a given model compared with the model's performance on the \textbf{Control} set. We can see that StainFuser is competitive with NST and significantly outperform both Ruifrok and Vahadane normalisation. With Ruifrok and Vahadane at best a given model performs on par with the un-normalised data and more often than not performs worse. Vahadane in particular with pale images, in the bottom row, and other outliers results in significantly worse performance. By contrast, NST and StainFuser improve every model's performance, particularly PathAI and Bern, for a multitude of targets. With the Bern model we see StainFuser improves performance for almost all sampled targets reinforcing prior findings that Bern is most robust to variations in color and compression \cite{vu2022nuclear}.

With PathAI and StarDist, we observe NST and StainFuser have very different patterns. By cross-referencing \Cref{fig:heatmap} against \Cref{fig:qualOverallMacro}, we observe that for target images with low saturation (lower region on the y-axis), PathAI performs better with StainFuser whereas it is the opposite with StarDist.

It is unknown to us what leads to such a significant divergence in the performance patterns. However, we speculate that different training augmentation regimes, while vastly increasing the original training data, also inadvertently and intractably diverge the data distribution observed by the models. Together with the inherent capacity of each model architecture in capturing such distribution, we ended up with each final model having widely different color foci for good performance. 


\subsection{Ablation Studies}

We study how various components of the training strategy affect final performance and explore the tradeoff between the number of denoising steps and generated image quality at inference time. For all downstream analysis in our ablations, we use PathAI's model.

\subsubsection{Number of Denoising steps}
\begin{table*}[t!]
\caption{Effect of denoising step number. Rows show performance across entire 101 target sets, $m\mathcal{PQ^+} AUC$ is using PathAI model on the CoNIC test sets. Image quality improves rapidly from 5 to 10 denoising steps and then starts to plateau.}
\label{tab:ablNSteps}
\centering
\small
\resizebox{0.75\textwidth}{!}{
\begin{NiceTabular}{cccccc}[]
\toprule
\RowStyle{\bfseries}
\Block[c]{}{Steps} & \lvl \Block[c]{}{Inference time (s)} & \Block[c]{}{$m\mathcal{PQ^+} AUC (\uparrow)$ } & \Block[c]{}{FID ($\downarrow$)} & \Block[c]{}{PSNR ($\uparrow$)} & \Block[c]{}{SSIM ($\uparrow$)} \\
\midrule
5 &\lvl {\color[HTML]{0000FF}{0.146 $\pm$ 0.002}} &\lvl 0.213 $\pm$ 0.011 &\lvl 42.150 $\pm$ 12.086 &\lvl 21.298 $\pm$ 0.704 &\lvl 0.853 $\pm$ 0.017\\
10 &\lvl 0.249 $\pm$ 0.003 &\lvl {\color[HTML]{0000FF}{0.214 $\pm$ 0.011}} &\lvl 34.393 $\pm$ 10.680 &\lvl 23.148 $\pm$ 0.965 &\lvl 0.865 $\pm$ 0.017\\
20 &\lvl 0.408 $\pm$ 0.002 &\lvl {\color[HTML]{0000FF}{0.214 $\pm$ 0.011}} &\lvl 32.760 $\pm$ 10.398  &\lvl 23.777 $\pm$ 1.090 &\lvl {\color[HTML]{0000FF}{0.868 $\pm$ 0.017}}\\
50 &\lvl 0.880 $\pm$ 0.003 &\lvl {\color[HTML]{0000FF}{0.214 $\pm$ 0.011}} &\lvl 32.211 $\pm$ 10.292 &\lvl 24.034 $\pm$ 1.151 &\lvl 0.868 $\pm$ 0.018\\
100 &\lvl 1.713 $\pm$ 0.237 &\lvl 0.213 $\pm$ 0.011 &\lvl {\color[HTML]{0000FF}{32.058 $\pm$ 10.268}} &\lvl {\color[HTML]{0000FF}{24.110 $\pm$ 1.166}} &\lvl 0.868 $\pm$ 0.018 \\
\bottomrule
\end{NiceTabular}
}
\end{table*}

\begin{table*}[t!]
\caption{Results for different image resolutions during training and inference. We train 2 models, 1 on $256^2$ data and 1 on $512^2$ data and then apply each on $256^2$ and $512^2$ unseen data observing that the larger resolution of $512^2$ data leads to better performance even when the model was trained on $256^2$ images. Best performance is highlighted in {\color[HTML]{0000FF}{blue}}.}
\label{tab:ablImgRes}
\centering
\small
\resizebox{0.75\textwidth}{!}{
\begin{NiceTabular}{cccccc}[]
\toprule
\RowStyle{\bfseries}
\Block{1-2}{Resolution} & & & &\\
\RowStyle{\bfseries}
\Block[c]{}{Training} & \lvl \Block[c]{}{Inference} & \Block[c]{}{$m\mathcal{PQ^+} AUC (\uparrow)$ } & \Block[c]{}{FID ($\downarrow$)} & \Block[c]{}{PSNR ($\uparrow$)} & \Block[c]{}{SSIM ($\uparrow$)} \\
\midrule
$256^2$ & $256^2$ &\lvl 0.100 $\pm$ 0.003 &\lvl 79.270 $\pm$ \phantom{1}7.238 &\lvl 19.827 $\pm$ 0.999 &\lvl 0.607 $\pm$ 0.008 \\
$512^2$ & $256^2$ &\lvl 0.117 $\pm$ 0.004 &\lvl 59.328 $\pm$ \phantom{1}6.663 &\lvl 20.014 $\pm$ 0.941 &\lvl 0.612 $\pm$ 0.009\\
$256^2$ & $512^2$ &\lvl 0.157 $\pm$ 0.006 &\lvl 40.164 $\pm$ \phantom{1}7.408 &\lvl 21.814 $\pm$ 1.036 &\lvl 0.808 $\pm$ 0.013\\
$512^2$ & $512^2$ &\lvl {\color[HTML]{0000FF}{0.214 $\pm$ 0.011}} &\lvl {\color[HTML]{0000FF}{32.760 $\pm$ 10.398}} &\lvl {\color[HTML]{0000FF}{23.777 $\pm$ 1.090}} &\lvl {\color[HTML]{0000FF}{0.868 $\pm$ 0.017}} \\
\bottomrule
\end{NiceTabular}
}
\end{table*}

\begin{table*}[t!]
\caption{Effect of image magnification during training on generated image quality and downstream performance. We trained 3 different models each with all 512 target sets at $512^2$ resolution. Best results are highlighted in {\color[HTML]{0000FF}{blue}}.}
\label{tab:ablImgMag}
\centering
\small
\resizebox{0.75\textwidth}{!}{
\begin{NiceTabular}{ccccc}[]
\toprule
\RowStyle{\bfseries}
\Block[c]{}{Magnification} &\lvl \Block[c]{}{$m\mathcal{PQ^+} AUC (\uparrow$) } & \Block[c]{}{FID ($\downarrow$)} & \Block[c]{}{PSNR ($\uparrow$)} & \Block[c]{}{SSIM ($\uparrow$)} \\
\midrule
20x &\lvl 0.214 $\pm$ 0.011 &\lvl 32.760 $\pm$ 10.398 &\lvl 23.777 $\pm$ 1.090 &\lvl 0.868 $\pm$ 0.017\\
40x &\lvl 0.209 $\pm$ 0.008 &\lvl 28.878 $\pm$ \phantom{1}7.718 &\lvl 22.585 $\pm$ 0.815 &\lvl 0.836 $\pm$ 0.008\\
20x \& 40x &\lvl {\color[HTML]{0000FF}{0.215 $\pm$ 0.007}} &\lvl {\color[HTML]{0000FF}{25.882 $\pm$ \phantom{1}8.233}} &\lvl {\color[HTML]{0000FF}{23.911 $\pm$ 0.816}} &\lvl {\color[HTML]{0000FF}{0.875 $\pm$ 0.010}} \\
\bottomrule
\end{NiceTabular}
}
\end{table*}

\begin{table*}[t!]
\caption{Comparison between StainFuser models trained with different volumes of data. Models are trained for 3 epochs at $512^2$ resolution, at 20x magnification. $m\mathcal{PQ^+} AUC$ results are using the PathAI model}
\label{tab:ablDataVol}
\centering
\small
\resizebox{0.75\textwidth}{!}{
\begin{NiceTabular}{ccccc}[]
\toprule
\RowStyle{\bfseries}
\Block[c]{}{Target Sets} &\lvl \Block[c]{}{$m\mathcal{PQ^+} AUC (\uparrow$) } & \Block[c]{}{FID ($\downarrow$)} & \Block[c]{}{PSNR ($\uparrow$)} & \Block[c]{}{SSIM ($\uparrow$)} \\
\midrule
64 &\lvl 0.168 $\pm$ 0.005 &\lvl 44.951 $\pm$ \phantom{1}8.836 &\lvl 19.304 $\pm$ 0.576 &\lvl 0.797 $\pm$ 0.010\\
128 &\lvl 0.204 $\pm$ 0.006 &\lvl 39.541 $\pm$ 11.062 &\lvl 21.648 $\pm$ 0.901 &\lvl 0.821 $\pm$ 0.010\\
256 &\lvl 0.212 $\pm$ 0.008 &\lvl 36.391 $\pm$ 10.602 &\lvl 21.836 $\pm$ 0.976 &\lvl 0.827 $\pm$ 0.014\\
512 &\lvl {\color[HTML]{0000FF}{0.214 $\pm$ 0.011}} &\lvl {\color[HTML]{0000FF}{32.760 $\pm$ 10.398}} &\lvl {\color[HTML]{0000FF}{23.777 $\pm$ 1.090}} &\lvl {\color[HTML]{0000FF}{0.868 $\pm$ 0.017}} \\
\bottomrule
\end{NiceTabular}
}
\end{table*}
We perform inference across all test sets with different numbers of denoising steps and analyse the impact this has on image quality and downstream model performance and report the results in \Cref{tab:ablNSteps}. Here, we see FID, PSNR and SSIM change by -10.092 FID, +2.812 PSNR and +0.015 SSIM between 5 and 100 denoising steps. However, there are diminishing returns when the number of denoising steps increases beyond 10. Due to the results of this ablation, we used 20 denoising steps for all other downstream analyses as this represented the best compromise between inference time, image quality and downstream performance.


\subsubsection{Importance of Image Resolution.}

We study the impact of image resolution by training StainFuser on two different image resolutions $512^2$ and $256^2$. These resolutions are the two most common resolutions for inference in CPath WSI-level work and thus allow us to explore whether a higher resolution is required for good performance. We utilized the PathAI model for evaluating the impacts of the resolution on the downstream task.

The quantitative results are reported in \Cref{tab:ablImgRes} and shown qualitatively in \Cref{fig:ablImgRes} of the supplementary material. From these results, we find that the higher resolution of $512^2$ images is crucial for both image quality and downstream performance. The StainFuser model trained on $512^2$ images drastically outperforms the model trained on $256^2$ images whether applied on $256^2$ or $512^2$ as shown in \Cref{tab:ablImgRes}. Furthermore, the $246^2$ trained model's performance improves when applied on $512^2$ images both in terms of image quality and downstream performance.

We hypothesize this is likely due to the frozen VAE we use in our architecture. This VAE was originally trained on $512^2$ images and as such likely has learned feature embeddings for pixel arrangements only found in images of this resolution or larger. As such when it is used to embed smaller images the embeddings do not contain sufficiently high-quality information for StainFuser to learn and apply the style transfer effectively.

Lastly, the improved performance of StainFuser at $512^2$ compared to $256^2$ also has positive connotations for downstream application at the WSI level as by normalizing at this resolution the number of tiles in a WSI that need to be processed is reduced by a factor of 4 providing significant computational speedup.

We use models trained on $512^2$ images for all other ablations due to the difference in downstream performance observed here.

\begin{figure*}[!t]
\centering
\includegraphics[width=0.9\textwidth]{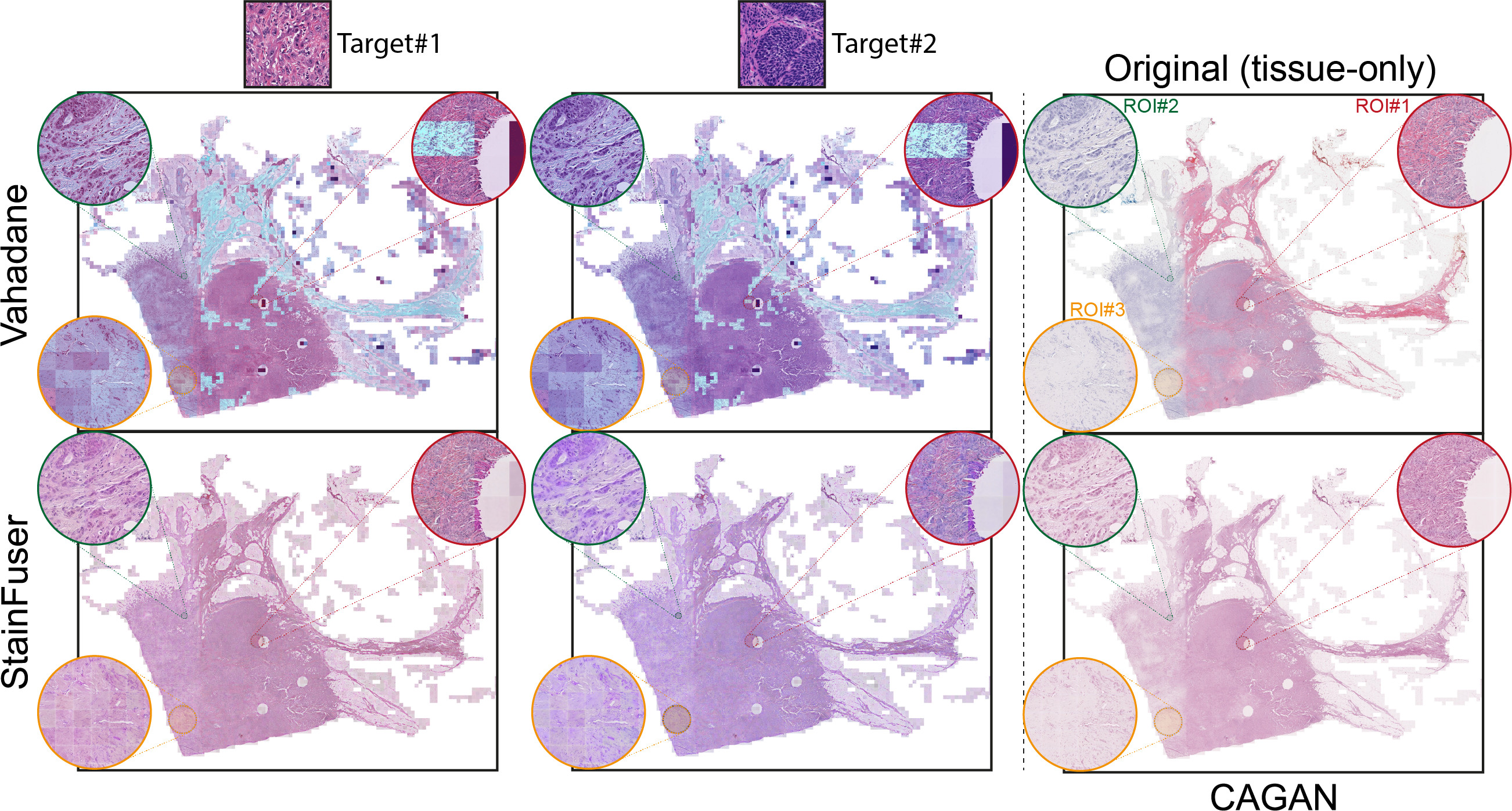}
\caption{
WSI inference comparison between Vahadane, StainFuser and CAGAN. The slide was chosen from a different anatomic site (\ie breast) and 2 target images were chosen from 2 different slides previously unseen by the StainFuser. 
}
\label{fig:qualWSI}
\end{figure*}

\subsubsection{Image Magnification}

We study the impact of image magnification by training StainFuser models on images with magnifications of 20x, 40x and a mixture of 20x and 40x during training. Similarly, we utilized PathAI model for evaluating downstream performance. Our results are included in \Cref{tab:ablImgMag}.

In the mixed image setting when a sample is fetched from the dataloader we randomly select a sample with probability 0.5, either a 40x or a 20x version of the same image, target pair at the given fixed image resolution \ie $512^2$. We find that models trained on 20x data marginally outperform those trained on 40x data in terms of PSNR and SSIM but not in FID.

In terms of downstream performance, the PathAI model performed marginally better with the 20x StainFuser data (+0.005 $m\mathcal{PQ^+} AUC$) compared to the 40x StainFuser data across the normalized test sets. Additionally, the StainFuser trained using 20x and 40x data outperforms both other training settings in terms of image quality, across FID, PSNR and SSIM, and in terms of downstream $m\mathcal{PQ^+} AUC$. This is potentially due to the distribution of image magnifications within the CoNIC test set where the majority of the images were captured at 20x magnification. By extension, the 20x and 40x model benefits by seeing all the magnifications within the testing set and our results show this is both in terms of image quality and downstream performance. Given this, we expect the 20x and 40x model to generalize better than the other models to other downstream tasks having been exposed to both magnifications in training.

\subsubsection{Data Volume}
We train 4 different StainFusers using a different number of target sets to explore the importance of the amount of our sampled training data on performance. Similarly, we utilized the PathAI model for evaluating downstream performance. Specifically, we use 64, 128, 256 and 512 target sets for our ablations. The target sets are chosen by sampling from the color distribution of the reference image of the given target set to encompass as much of the overall color space as possible. Furthermore, as the number of target sets increases the higher number always includes all of the previous target sets. \ie the target sets in the 128 experiment contain all the target sets of the 64 experiment in addition to 64 others \etc. These target set numbers correspond to \num{262144}, \num{524288}, \num{1048576} and \num{2097152} unique images in each training set respectively. 

We report the results in \Cref{tab:ablDataVol}. Here, we observe that the more data used for training the higher quality images StainFuser generates on unseen data and the better the PathAI model performs on the corresponding normalized test datasets.
Here, we see the performance improvement is much steeper when going from 64 to 128 target sets. However, beyond this point the improvements slowly plateau.

On the whole, it is clear that unsurprisingly the more diverse data StainFuser is trained on the higher quality images it generates and the better downstream models using its normalised data perform.

\subsection{Inference on Whole Slide Images}

We further qualitatively compare StainFuser against Vahadane and CAGAN at WSI-level in \Cref{fig:qualWSI}. The model used for StainFuser was trained on the 512 target set using patches extracted at both $20\times$ and $40\times$ magnification level with image size of $512^2$. For inference, images were processed at $20\times$ magnification with image size of $512^2$, and StainFuser's denoising step was set to 20. We used TIAToolbox's\cite{pocock2022tiatoolbox} implementation for Vahadane method. Only tissue sections were processed.
As can be seen from \Cref{fig:qualWSI}, StainFuser shows a more consistent performance across the entire slide, whereas Vahadane's varies significantly. This can be observed in \textbf{ROI\#3} when using \textbf{Target\#1} and \textbf{Target\#2}. On the macro scale, Vahadane can also fail in certain regions and generate highly distorted images (\ie the blueish color patches), as shown in \textbf{ROI\#1}. Meanwhile, StainFuser's result generally appears less vibrant than Vahadane's, as demonstrated by \textbf{ROI\#2}. Despite this, from the same region, StainFuser still managed to achieve a clearer distinction between the tumor and the stroma compared to Vahadane. In comparison to Vahadane and StainFuser, although CAGAN transforms the input image only to one particular color domain as noted previously, the finer details of the image are significantly compromised. We illustrate this issue further in \Cref{fig:wsiInferenceQK} and \Cref{fig:wsiInferenceAA} of the Supplementary Material.

\subsection{Limitations}


While StainFuser generates high-quality images with clear contrast between important tissue components, we identify the following limitations in our work. First, the Stable Diffusion backbone is GPU memory intensive for training while inference is not, requiring less than 16GB of RAM using a batch size of 4 images of size $512^2$. Furthermore, the data curation we explored, while comprehensive, is restricted to three organs and does not represent the entire spectrum of tissue staining and morphologies possible. Additionally, curating data using NST is expensive, costing us over 10 thousand GPU hours. Finally, our method can sometimes produce slightly desaturated, less vibrant images compared to other approaches. While we've demonstrated this does not lead to worse image quality or downstream performance it is unclear what is causing this qualitative defect. 

\section{Conclusion}
\label{sec:conclusion}

\noindent
We present StainFuser, a novel method for stain normalization based on conditional diffusion models. For our approach, we curated to our knowledge the first, large-scale stain normalization dataset of over two million images. When trained on this dataset StainFuser achieves superior results compared to existing handcrafted and GAN-based methods in terms of image quality and downstream performance on the challenging CoNIC dataset while being 30 times faster than the current SoTA neural style transfer method. In addition, StainFuser achieves substantially better performance when used for WSI inference with superior color consistency between adjacent tiles and variations in stain compared to other methods. We believe our work provides a different perspective on the stain normalization task and the application of diffusion models in CPath.

\section*{CRediT authorship contribution statement}
\textbf{Robert Jewsbury:} Conceptualization, Data curation, Formal analysis, Investigation, Methodology, Project administration, Software, Validation, Visualization, Writing – original
Writing – review \& editing.
\textbf{Ruoyu Wang:} Data curation, Formal analysis, Investigation, Methodology, Software, Validation, Visualization, Writing – original draft, Writing – review \& editing.
\textbf{Abhir Bhalerao:} Methodology, Resources, Writing – review \& editing.
\textbf{Nasir Rajpoot:} Funding acquisition, Resources, Writing – review \& editing.
\textbf{Quoc Dang Vu:} Conceptualization, Methodology, Software, Supervision, Writing – original draft, Writing – review \& editing.

\section*{Data Availability}
The data used in this research is publicly available and the link to it is cited in the manuscript.

\section*{Acknowledgments}
RJ and NR report financial support from GlaxoSmithKline, United Kingdom, NR's support is outside of this work. RW reports funding from the General Charities of the City of Coventry and the Computer Science Doctoral Training Centre at the University of Warwick. NR reports financial support provided by UK Research and Innovation (UKRI). NR is a co-founder of Histofy Ltd. 

\bibliographystyle{IEEEtran}
\bibliography{main}

\appendix

\setcounter{subsection}{0}
\renewcommand{\thesubsection}{A\arabic{subsection}}
\setcounter{figure}{0}
\renewcommand{\thefigure}{A\arabic{figure}}
\setcounter{table}{0}
\renewcommand{\thetable}{A\arabic{table}}
\renewcommand{\theequation}{A\arabic{equation}}

\clearpage
\section*{Supplementary Material}
\subsection{Implementation details}
\label{sec:suppImpl}
We use a Stable Diffusion v2.1 model \cite{rombach2022ldm} pre-trained on LAION-5B \cite{schuhmann2022laionb} as the backbone for our model\footnote{Backbone pre-training details are available at the model card: \url{https://huggingface.co/stabilityai/stable-diffusion-2-1-base}}. We train StainFuser with AdamW \cite{loshchilov2018decoupled} with a learning rate of $1\mathrm{e}{-5}$ and weight decay of $1\mathrm{e}{-2}$ for 3 epochs with an effective batch size of 32. Training the full model on $512^2$ images with 512 target sets took 81 hours on 2 A100 GPUs with 16 images per GPU.

\begin{figure*}[h]
\centering
\includegraphics[width=\columnwidth]{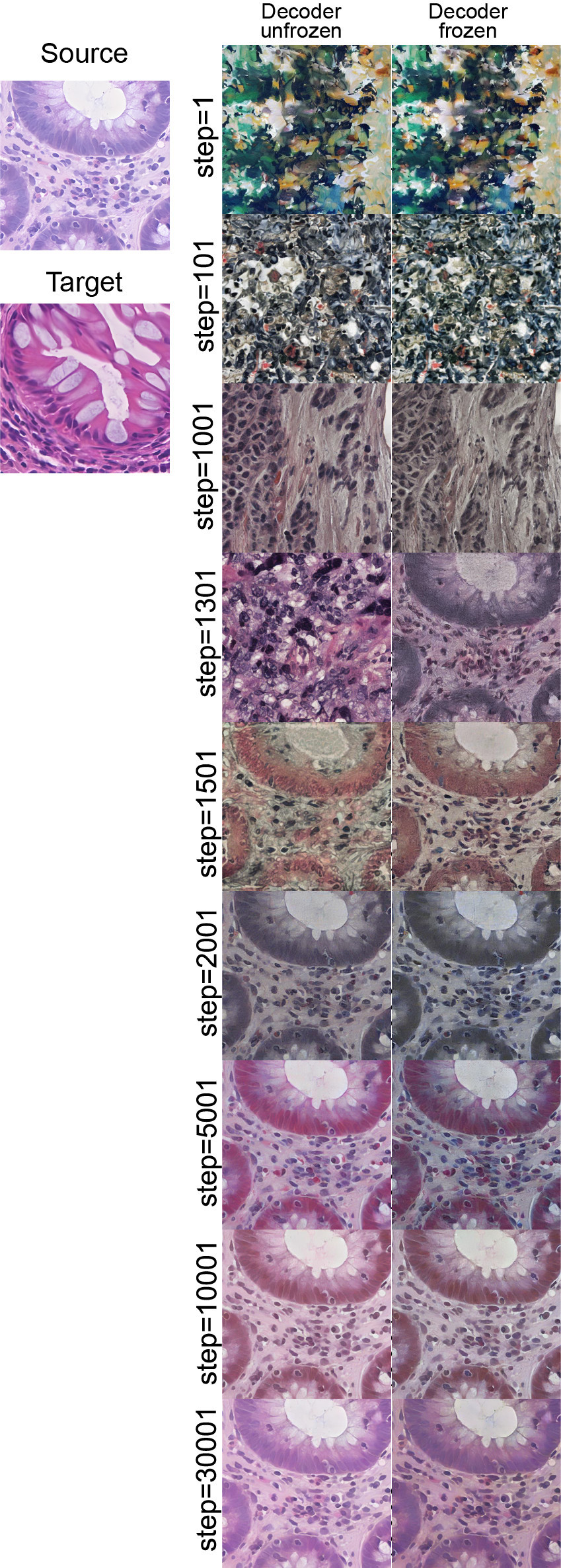}
\caption{Qualitative comparisons between decoder frozen and unfrozen during training shown in different optimization steps.
}
\label{fig:sudden_converg}
\end{figure*}

\subsubsection{Sudden convergence phenomenon}
In Zhang \etal \cite{ControlNet}, the authors reported a sudden convergence phenomenon during model training, which was also observed in our experiments. In the early training stage, the model can generate high-quality images with histological features, but they do not adhere to the guidance provided by the source image condition $p^s$. As shown in \Cref{fig:sudden_converg}, the model suddenly learned how to generate images based on the guidance from $p^s$ after a certain number of optimization steps.

\subsubsection{Decoder frozen vs. unfrozen during training}
A large-scale training strategy was also proposed in Zhang \etal \cite{ControlNet}, which involves initially training only the conditioning component of the model for a large number of steps, and the entire model, including the stable diffusion component, is then trained jointly. Given our constraints in computational resources, we explored whether unlocking only the decoder part of the stable diffusion component could enhance training speed and convergence. Therefore, we trained two models: one with the frozen decoder and the other with the unfrozen decoder. We then observed the performance of each model on an unseen validation set at various optimization steps.

We report the results in \Cref{fig:sudden_converg}. We can see that the sudden convergence phenomenon appeared earlier on the model with a frozen decoder. However, after both models learned how to generate the morphological content based on $p^s$, the model with the unfrozen decoder generates images with a better stain and image quality. We hypothesize that because the model with a frozen decoder only needs to optimize its conditioner, it (StainFuser) therefore learns faster adherence to the guiding signal from $p^s$. However, once the model adheres to the signal $p^s$ sufficiently, the one with a frozen decoder has trouble integrating the stain properties of $p^t$ into the final output, thus achieving less desirable image quality compared to the one with an unfrozen decoder.

\subsection{Additional Illustrative Results}
\label{sec:addResults}

\begin{figure*}[ht]
\centering
\includegraphics[width=0.85\paperwidth]{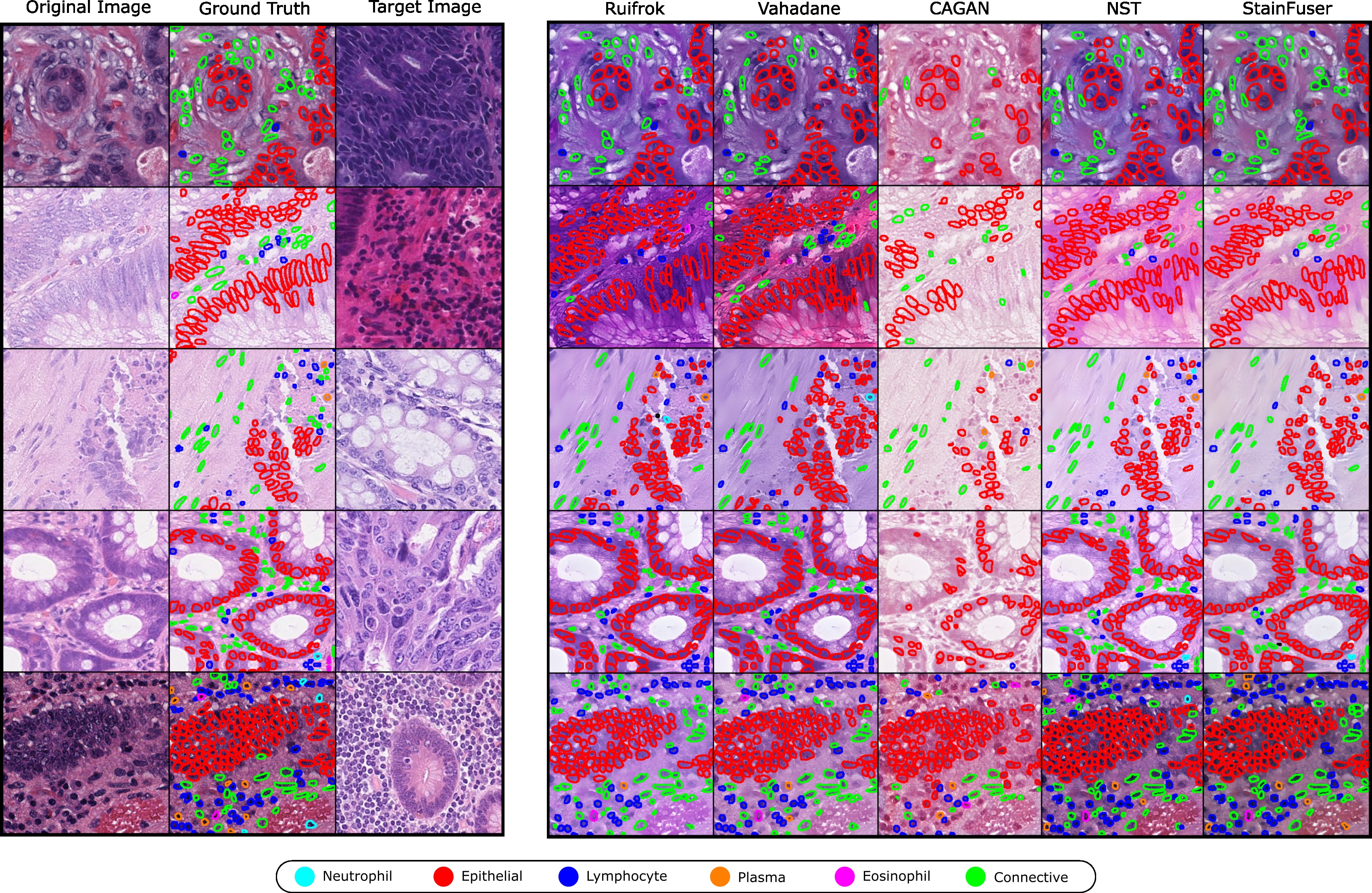}
\caption{Qualitative results of PathAI model applied on each normalization method
}
\label{fig:qualPathAI}
\end{figure*}
\clearpage

\begin{figure*}[ht]
\centering
\includegraphics[width=0.85\paperwidth]{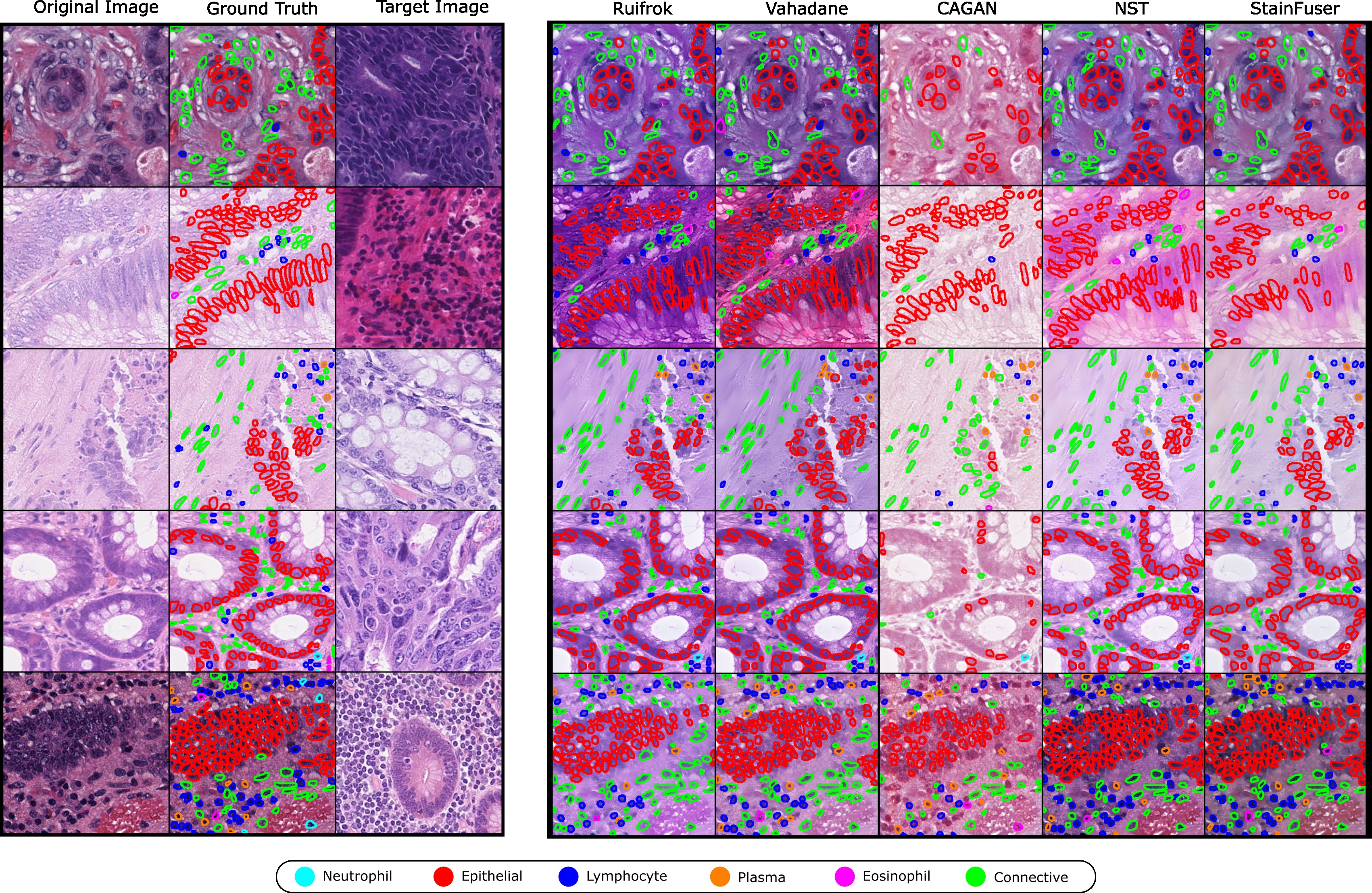}
\caption{Qualitative results of Bern model applied on each normalization method
}
\label{fig:qualBern}
\end{figure*}
\clearpage

\begin{figure*}[ht]
\centering
\includegraphics[width=0.85\paperwidth]{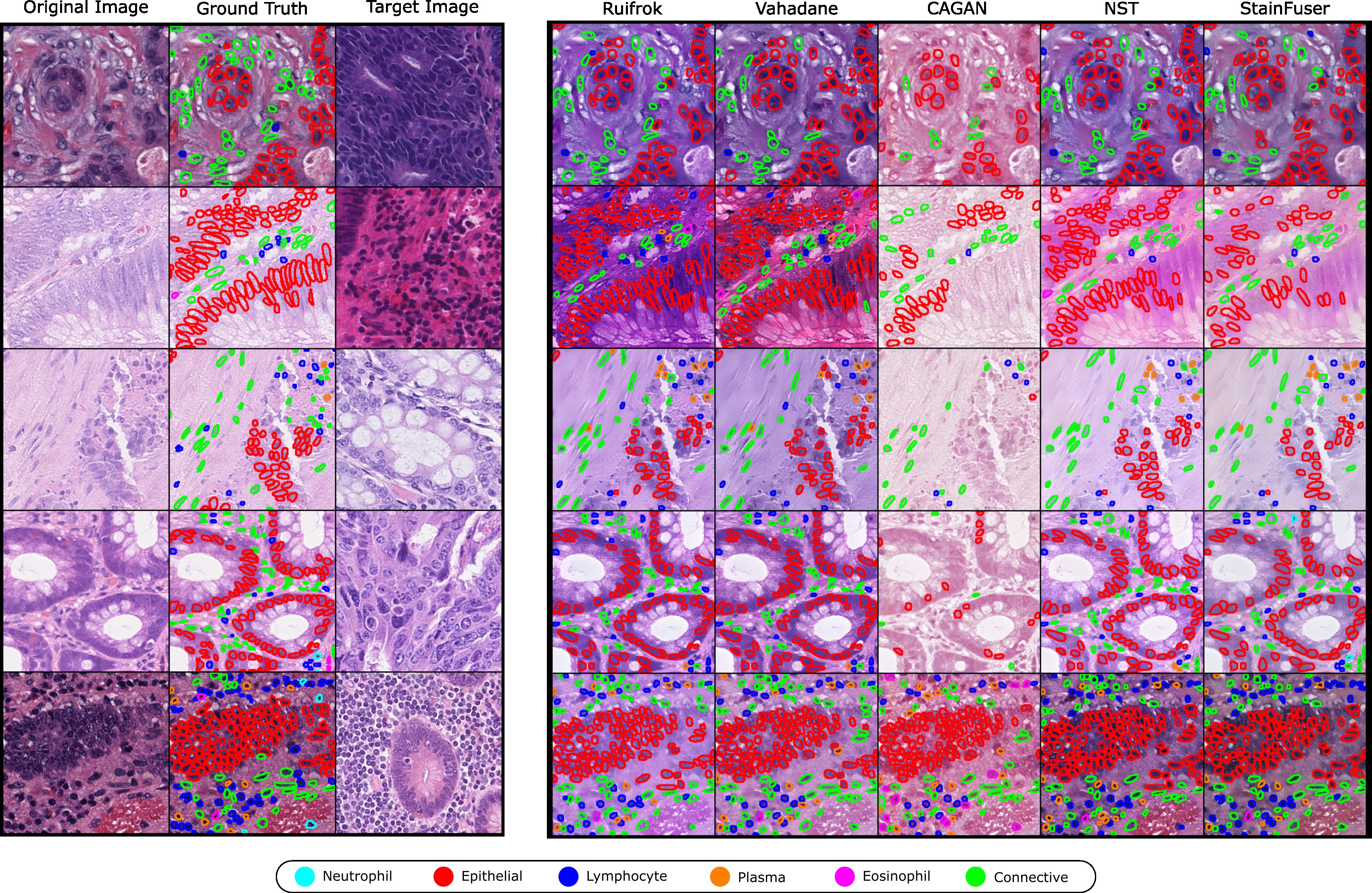}
\caption{Qualitative results of StarDist model applied on each normalization method
}
\label{fig:qualStarDist}
\end{figure*}
\clearpage

\begin{figure*}[h]
\centering
\includegraphics[width=1.00\textwidth]{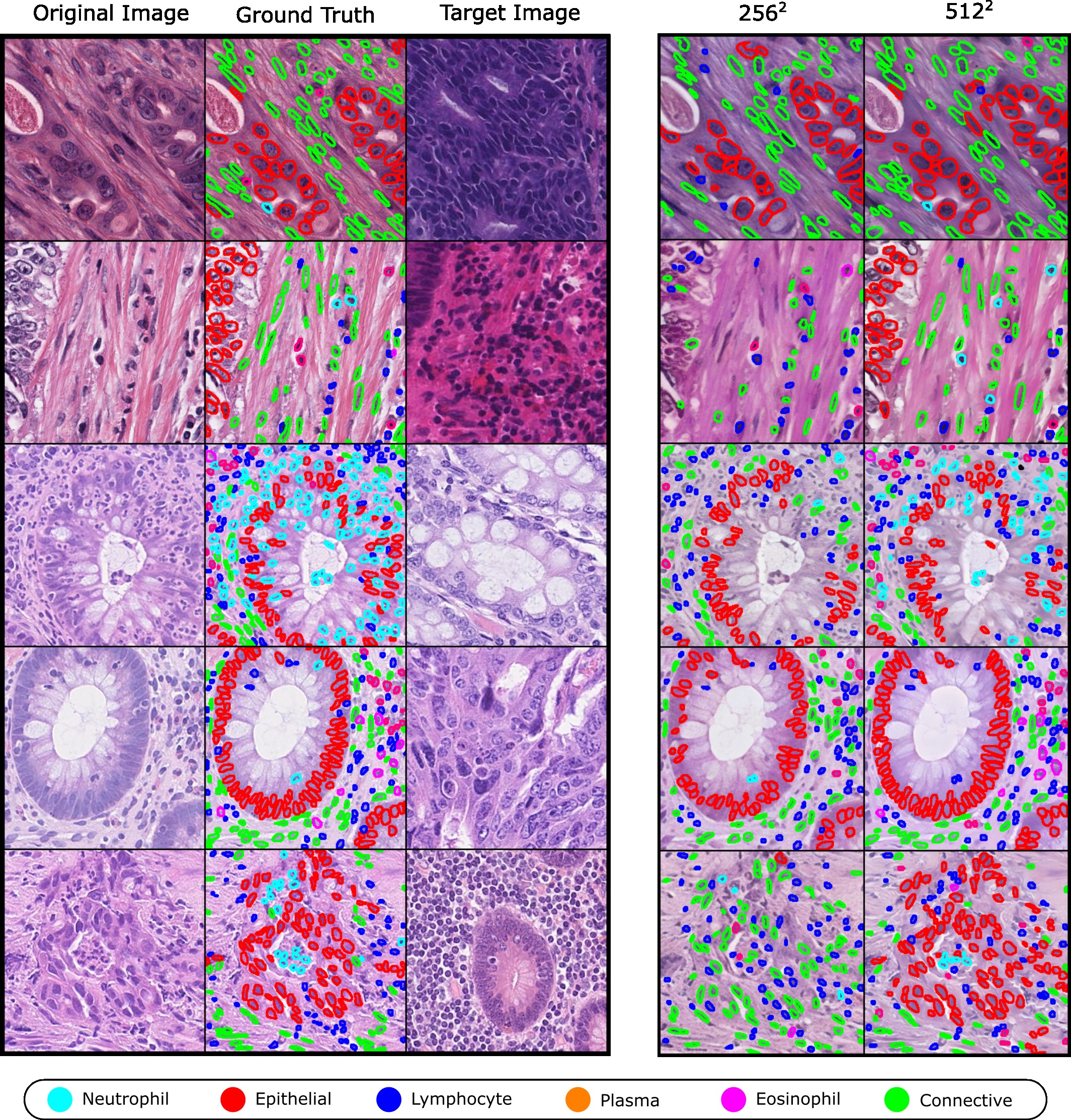}
\caption{Qualitative results showing the influence of image resolution for downstream performance. We observe that the normalized images generated by StainFuser trained at $256^2$ can lead to the misclassification of nuclei (bottom row) and missed nuclei (second and third row). All predictions are using the PathAI model.
}
\label{fig:ablImgRes}
\end{figure*}
\clearpage

\begin{figure*}
\centering
\includegraphics[width=1\textwidth]{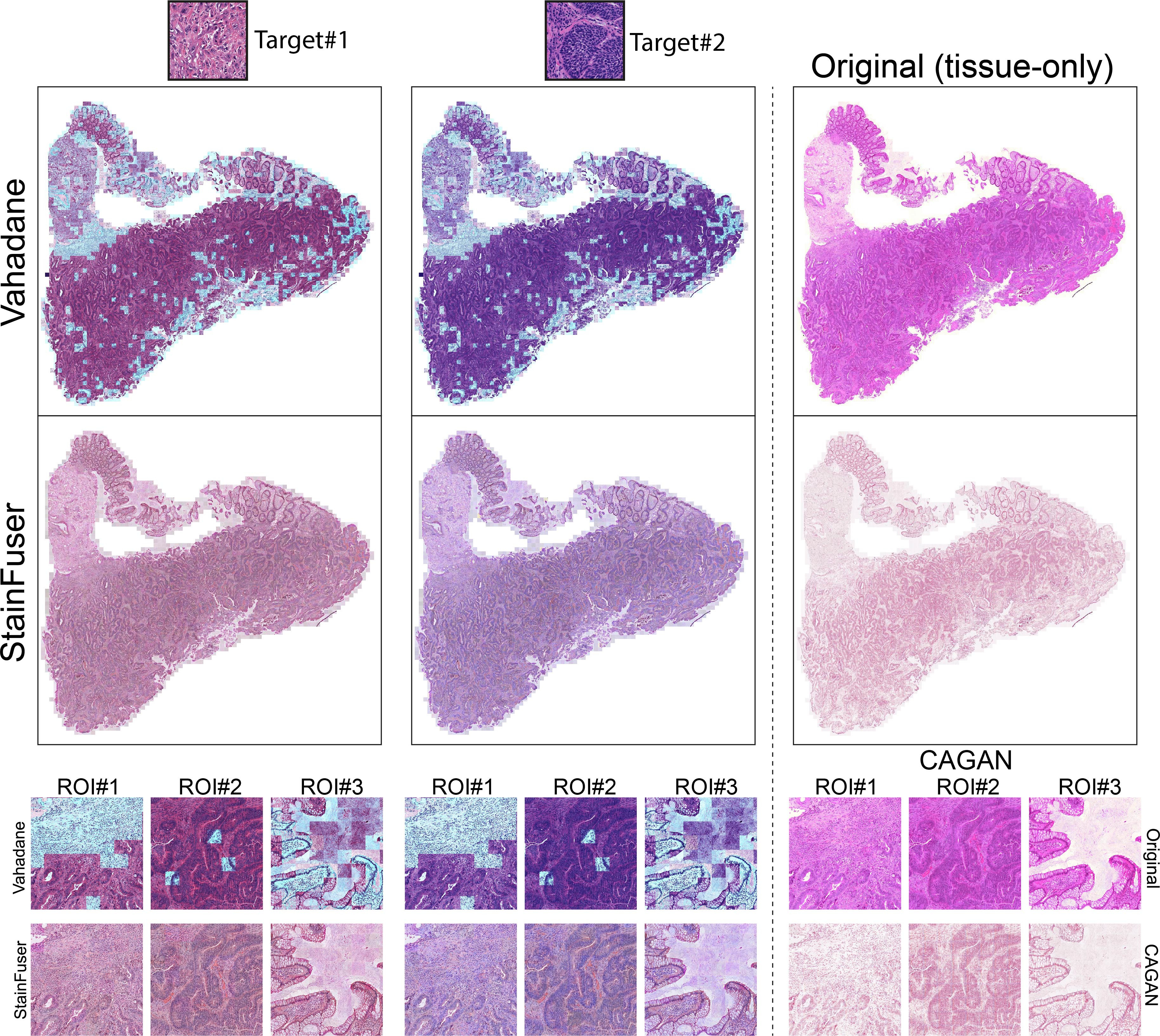}
\caption{
WSI inference comparison between Vahadane, StainFuser and CAGAN. The slide was chosen from TCGA-COAD and 2 target images were chosen from 2 different unseen slides to the StainFuser. 
}
\label{fig:wsiInferenceAA}
\end{figure*}
\clearpage

\begin{figure*}
\centering
\includegraphics[width=1\textwidth]{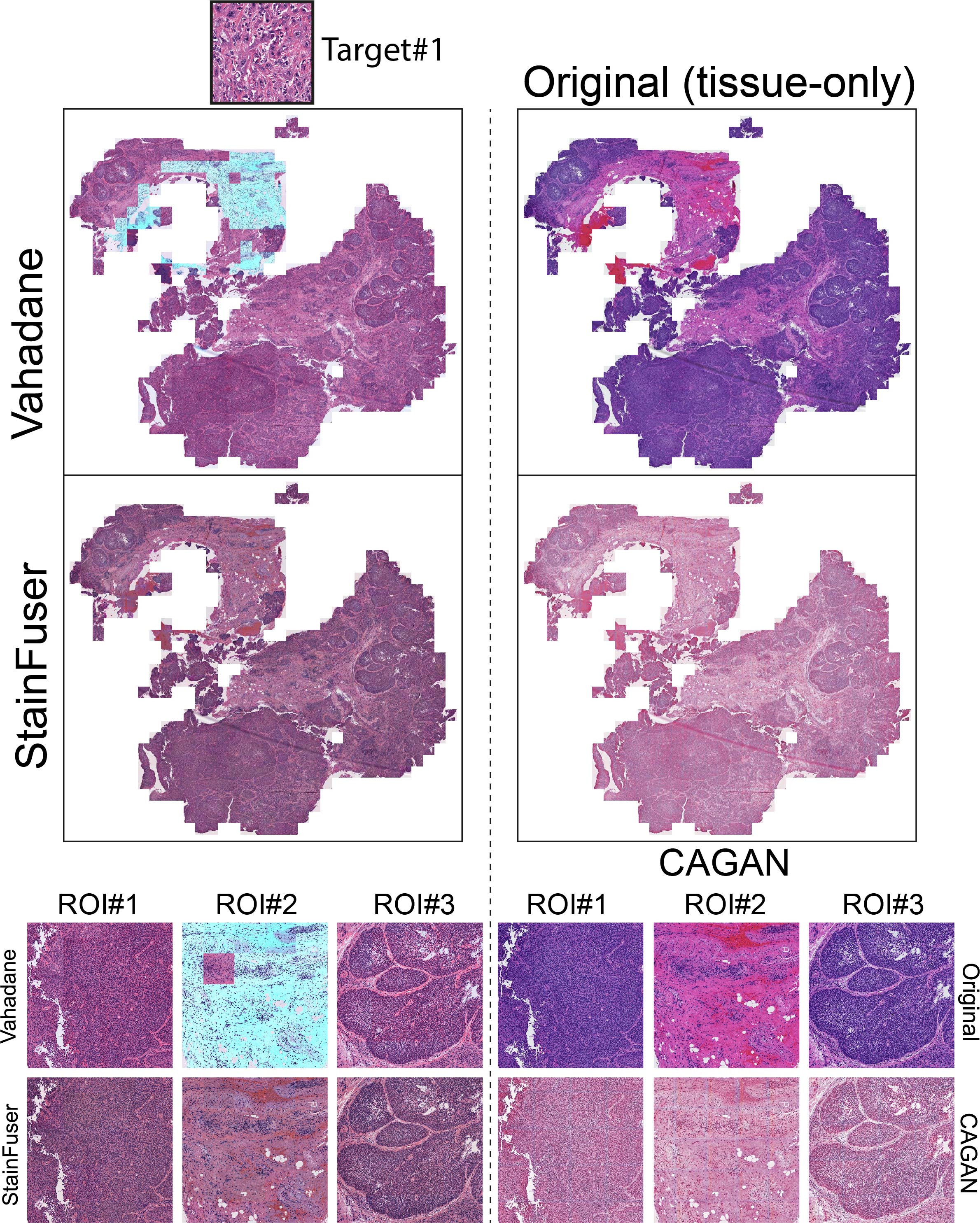}
\caption{
WSI inference comparison between Vahadane, StainFuser and CAGAN. The slide was chosen from TCGA-HNSC and 1 target image was chosen from 1 unseen slide to the StainFuser. 
}
\label{fig:wsiInferenceQK}
\end{figure*}
\clearpage

\begin{figure*}
    \centering
    \includegraphics[width=0.875\paperwidth]{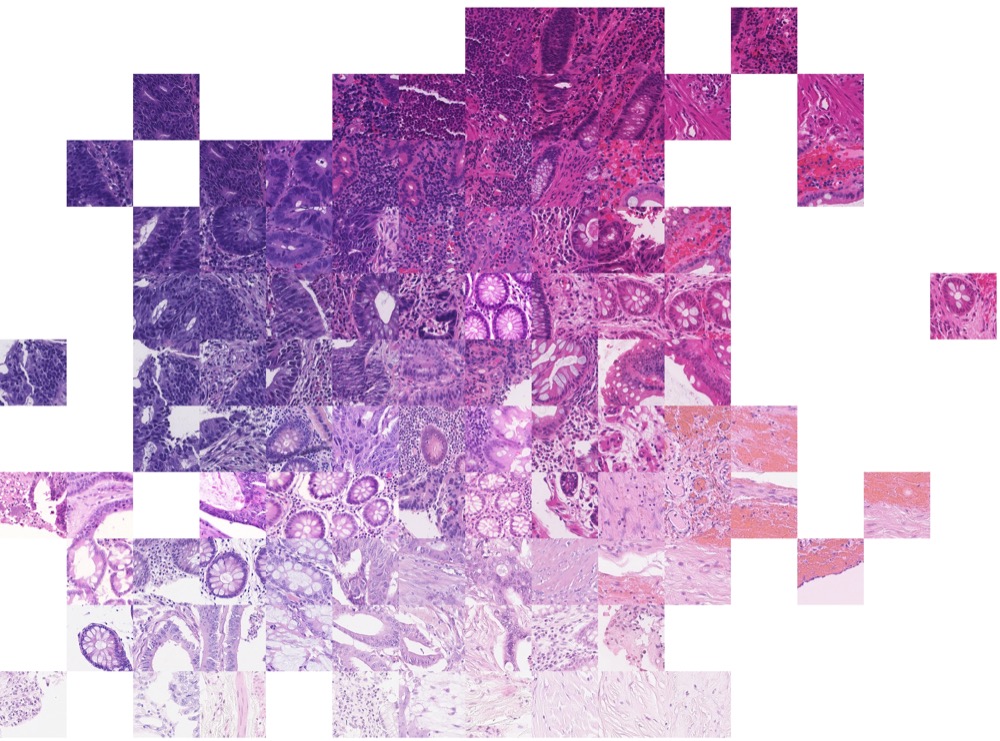}
    \caption{High-resolution version of sampled references}
    \label{fig:highResRefs}
\end{figure*}
\clearpage

\begin{figure*}
    \centering
    \includegraphics[width=0.875\paperwidth]{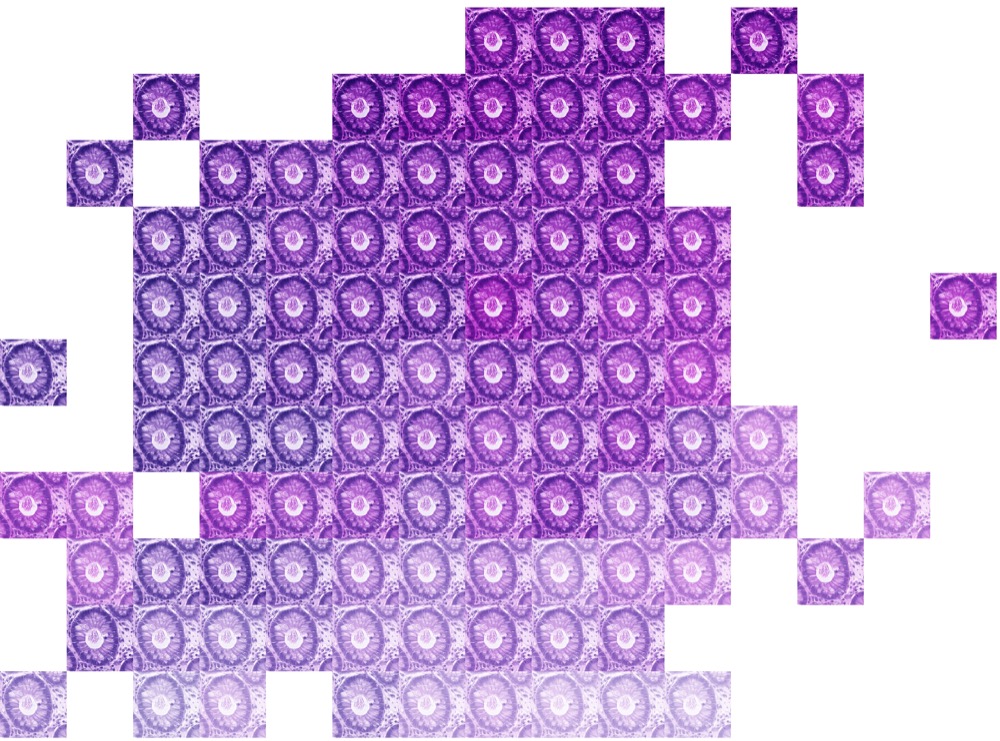}
    \caption{High-resolution version of image normalised by Ruifrok \cite{ruifrok2001} with respect to references in \Cref{fig:highResRefs}}
    \label{fig:highResRefRui}
\end{figure*}
\clearpage

\begin{figure*}
    \centering
    \includegraphics[width=0.875\paperwidth]{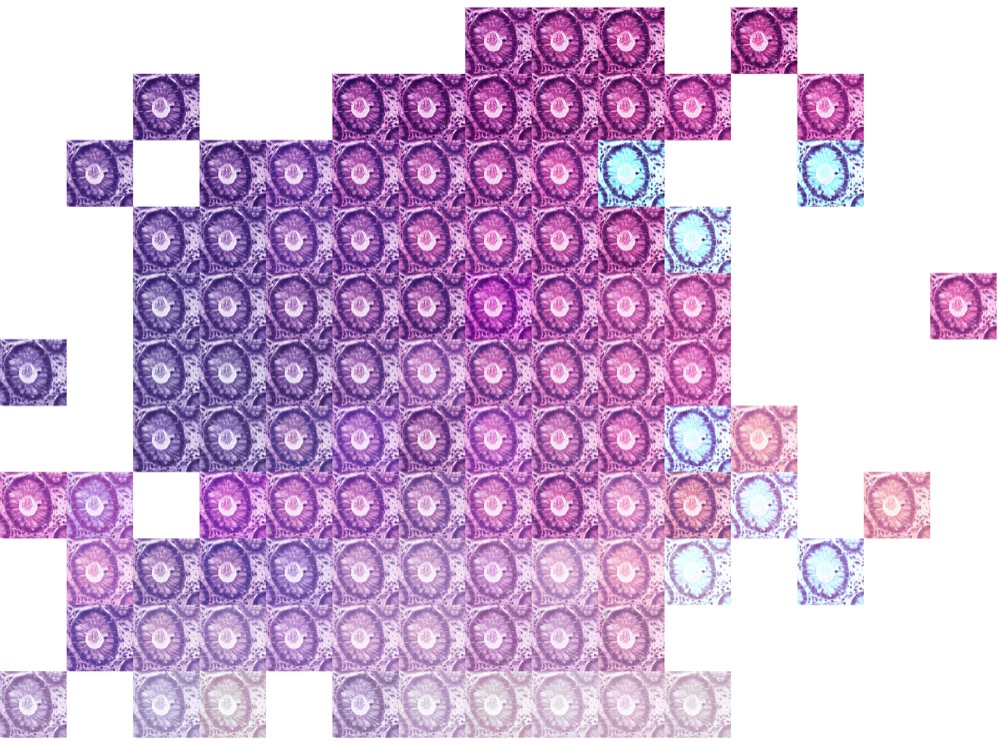}
    \caption{High-resolution version of image normalised by Vahadane \cite{vahadane2016structure} with respect to references in \Cref{fig:highResRefs}}
    \label{fig:highResRefVAh}
\end{figure*}
\clearpage

\begin{figure*}
    \centering
    \includegraphics[width=0.875\paperwidth]{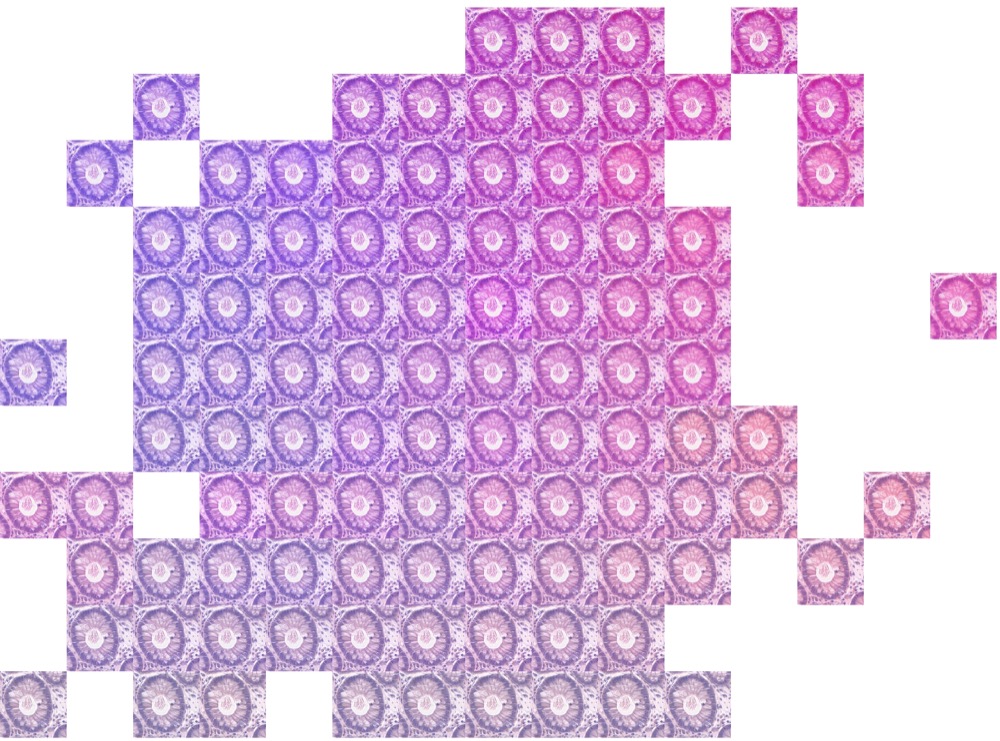}
    \caption{High-resolution version of image normalised by NST \cite{neural_style_transfer} with respect to references in \Cref{fig:highResRefs}}
    \label{fig:highResRefNST}
\end{figure*}
\clearpage

\begin{figure*}
    \centering
    \includegraphics[width=0.875\paperwidth]{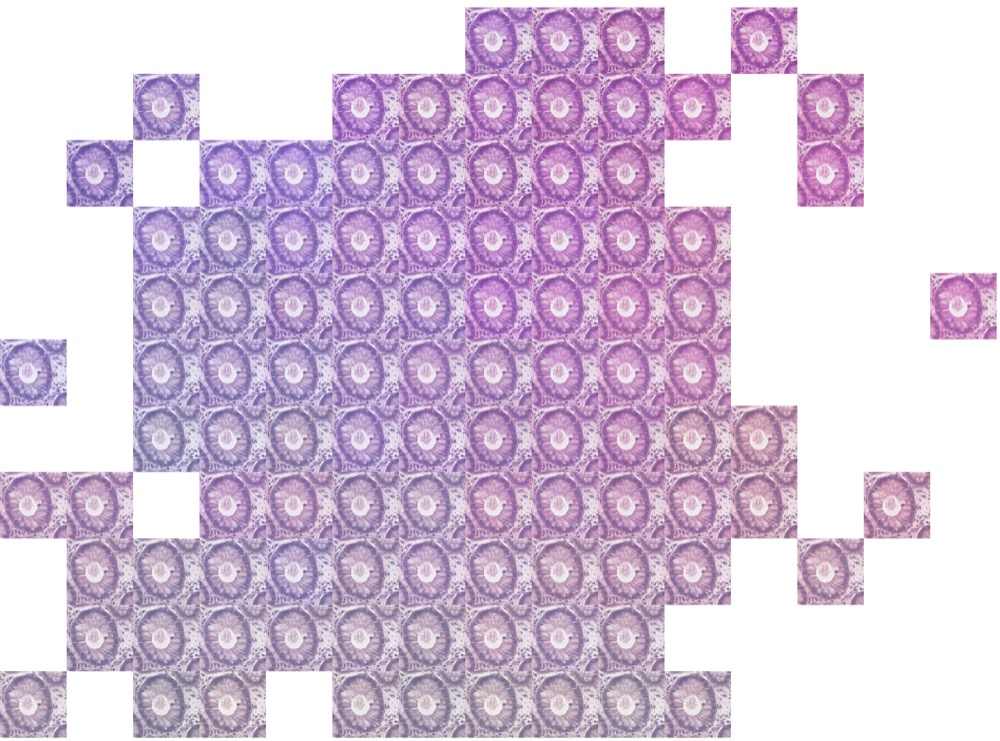}
    \caption{High-resolution version of image normalized by StainFuser with respect to references in \Cref{fig:highResRefs}}
    \label{fig:highResRefSF}
\end{figure*}

\end{document}